\documentclass[twocolumn,prl,aps,floatfix,nopacs,superscriptaddress]{revtex4}
\usepackage{float}
\usepackage{xcolor}
\usepackage{graphicx}
\usepackage{hyperref}
\usepackage{amsmath}
\hypersetup{backref,pdfpagemode=FullScreen,colorlinks=true}
\hypersetup{
  colorlinks=ture,
  linkcolor=blue,
  urlcolor=blue,
  citecolor=blue,
  }

\begin{document}

\title{Precompression engineering of metal-insulator transition and magnetism in designed breathing kagome systems}
\author{Qingzhuo Duan}
\affiliation{School of Physics and Astronomy, and Key Laboratory of Multiscale Spin Physics (Ministry of Education), Beijing Normal University, Beijing 100875, China}

\author{Hongdao Zhuge}
\affiliation{School of Physics and Astronomy, and Key Laboratory of Multiscale Spin Physics (Ministry of Education), Beijing Normal University, Beijing 100875, China}

\author{Ying Liang}
\email{liangying@hebtu.edu.cn}
\affiliation{College of Physics, Hebei Normal University, and Hebei Advanced Thin Films Laboratory, Shijiazhuang 050024, China}
\affiliation{School of Physics and Astronomy, and Key Laboratory of Multiscale Spin Physics (Ministry of Education), Beijing Normal University, Beijing 100875, China}

\author{Tianxing Ma}
\email{txma@bnu.edu.cn}
\affiliation{School of Physics and Astronomy, and Key Laboratory of Multiscale Spin Physics (Ministry of Education), Beijing Normal University, Beijing 100875, China}
\date{\today}

\begin{abstract}
Kagome materials featuring dispersive Dirac cones and topological flat bands exhibit unique electronic and magnetic properties. 
However, kagome compounds with tunable electrical conductivity remain scarce, which severely impedes their device applications. 
Here, based on density functional theory (DFT) and Boltzmann transport theory, we introduce the breathing effect into kagome materials $\mathrm{Nb_3XCl_7}$ (X = F, Cl, Br, I) via chemical precompression, thereby inducing a metal-insulator transition and magnetic variation. We determine that the band structures, optical absorption spectra and magnetic ground states agree well with experimental results at the effective correlation strength $U_{\text{eff}} = 2$ eV. The calculated conductivity and magnetic properties reveal that the monolayer $\mathrm{Nb_3Cl_8}$ and $\mathrm{Nb_3XCl_7}$ undergoes transitions from paramagnetic metals to Mott insulators at $U_{\text{eff}} = 1$ eV and $t_{\text{out}}/t_{\text{in}} = 0.6674$, respectively. Our detailed analysis establishes that the stronger breathing effect corresponds to enhanced chemical precompression, which reduces the region of free electron gas between intercell Nb atoms and facilitates the metal-insulator transition. Finally, we propose several viable synthesis routes for $\mathrm{Nb_3FCl_7}$, $\mathrm{Nb_3BrCl_7}$, and $\mathrm{Nb_3ICl_7}$, providing predictive guidance for experimental studies. Our study establishes a practical framework for investigating the breathing effect in correlated kagome systems and yields valuable insights into the mechanisms underlying metal-insulator transition and magnetic properties in real breathing kagome materials.
\end{abstract}

\maketitle

\section{I. INTRODUCTION}
\vspace{-0.2cm}

The kagome lattice, a geometrically frustrated network of corner-sharing triangles, initially attracted considerable research interest in the context of quantum spin liquids and has since emerged as an excellent platform for investigating electron correlation effects, such as metal-insulator transitions(MIT), magnetism, unconventional superconductivity, and nontrivial topological phenomena\cite{PhysRevLett.127.236401,PhysRevLett.134.086902,nature08917,Nature.11659,science.aab2120,PhysRevLett.126.247001,ZHU2023157817,FANG2024112725,Yang_2024}. This unique lattice geometry gives rise to two characteristic features in the electronic band structure: Dirac cones and flat bands. The Dirac cones originate from the honeycomb sublattice symmetry, analogous to that of graphene. When spin–orbit coupling and magnetism are introduced to break time-reversal symmetry, these Dirac cones at the high-symmetry points of the Brillouin zone can be gapped out, leading to nontrivial topological states\cite{Nature.s41586-020-2482-7,PhysRevLett.106.236802,nature25987}. However, flat bands arise from geometric frustration-induced destructive interference of wave functions, resulting in electronic states that are localized on the hexagonal rings. These highly degenerate flat bands have attracted significant attention due to the strongly correlated electronic phenomena they host, including quantum spin liquids, high-temperature superconductivity, and fractional quantum Hall effects\cite{PhysRevX.11.041010,Nature.s41586-021-03983-5,PhysRevLett.126.247001,PhysRevLett.106.236802,Nat.Mater.s41563-021-01034-y}. 

Representative materials such as $\rm AV_3Sb_5$ (A = K, Rb and Cs), $\mathrm{Fe_3Sn_2}$, $\mathrm{Co_3Sn_2S_2}$, $\mathrm{Ni_3In_2S_2}$, $\mathrm{CoSn}$, $\mathrm{FeGe}$, and $\mathrm{TbMn_6Sn_6}$ have been found to exhibit charge density waves and unconventional superconductivity with topological properties \cite{10.1093/nsr/nwac199,acs.nanolett.1c02270,adma.201806622,s41524-022-00838-z,s41467-020-17465-1,PhysRevLett.129.166401,Xu2022}. Of particular importance, the kagome material $\mathrm{Nb_3Cl_8}$ has been successfully synthesized \cite{acs.nanolett.2c00778}, wherein lattice distortions engender a breathing effect. The breathing kagome lattice breaks sixfold rotational symmetry, opening the Dirac cone and altering multiple band gaps. Subsequently, $\mathrm{Nb_3TeCl_7}$ was synthesized by solid-state reaction \cite{adma.202301790}. Experimental and theoretical results indicate that substituting Te atoms in $\mathrm{Nb_3TeCl_7}$ can effectively modulate the bandwidth and energy position of Nb sites. Interestingly, $\mathrm{Nb_3Br_8}$ crystallizing in the $R\overline{3}m$ space group exhibits a sign reversal in temperature-dependent conductivity upon compression to 24.8 GPa, signaling a pressure-driven MIT \cite{k3fv-9hss}. Most previously discovered kagome materials are gapless metals, a limitation that severely restricts their applications in logic and optoelectronic devices. Although several kagome materials have achieved semiconducting behavior with moderate band gaps \cite{acs.nanolett.2c00778,adma.202301790,k3fv-9hss}, the specific relationship between breathing strength and the MIT remains elusive, and the underlying physical mechanisms have yet to be fully elucidated.

Common kagome materials exhibit antiferromagnetic (AFM) order, such as FeGe, $\rm YMn_{6}Sn_{6}$, and $\rm CsCr_{3}Sb_{5}$ \cite{Nature.022.05034,PhysRevB.103.014416,Nat.Commun.16.1375}. Theoretically, density-matrix renormalization group calculations on the half-filled kagome lattice Hubbard model reveal that the system develops AFM order once the on-site Coulomb repulsion $U$ surpasses a critical value $U_{c3}$. This motivates the active exploration and control of exotic magnetic phases in kagome systems. $\mathrm{YbTi_3Bi_4}$ features a distorted kagome lattice and rare-earth zigzag chains \cite{CommunMater.024.00513}. By substituting rare-earth atoms in the zigzag chains, nonmagnetic (NM) $\mathrm{YbTi_3Bi_4}$ can be transformed into short-range-ordered $\mathrm{PrTi_3Bi_4}$ and eventually into ferromagnetic (FM) $\mathrm{NdTi_3Bi_4}$. Interestingly, $\mathrm{Ti_3I_8}$ with a breathing kagome lattice exhibits coexisting FM order \cite{PhysRevB.104.L060405}. 
The cluster units composed of Ti trimers constitute ``cluster magnets'' that exhibit a certain degree of electron localization\cite{Nakamura_2005,ic701011r}. This localization is stabilized by an appropriate balance between inter-cluster transfer integrals and on-site Coulomb repulsion. Consequently, cluster magnets emerging from the breathing effect are expected to give rise to exotic phenomena \cite{haraguchi2017}.

Aiming to clarify the interplay between breathing strength and the MIT and to explore emergent magnetic properties, we have conducted a systematic theoretical investigation of correlated kagome systems using the DQMC method \cite{cpl_42_9_090712}. Our results demonstrate that the interplay between the breathing effect and interaction strength drives the transition from a PM metal to a Mott insulator in the kagome system. Guided by these theoretical insights, we aim to achieve continuous tuning of the breathing effect in real materials, thereby exploring how the breathing effect governs MIT and magnetic properties in correlated kagome materials. Recently, a 2D layered kagome material $\mathrm{Nb_3Cl_8}$ and its doped counterpart $\mathrm{Nb_3TeCl_7}$ have been successfully synthesized via solid-state reaction and mechanical exfoliation, respectively \cite{acs.nanolett.2c00778,adma.202301790}. Their facile synthesis routes offer significant advantages for experimental studies. Therefore, we select the kagome material $\mathrm{Nb_3Cl_8}$ as the parent structure and substitute the Cl atom atop the Nb trimer with Group VIIA elements F, Br, and I (As shown in \hyperref[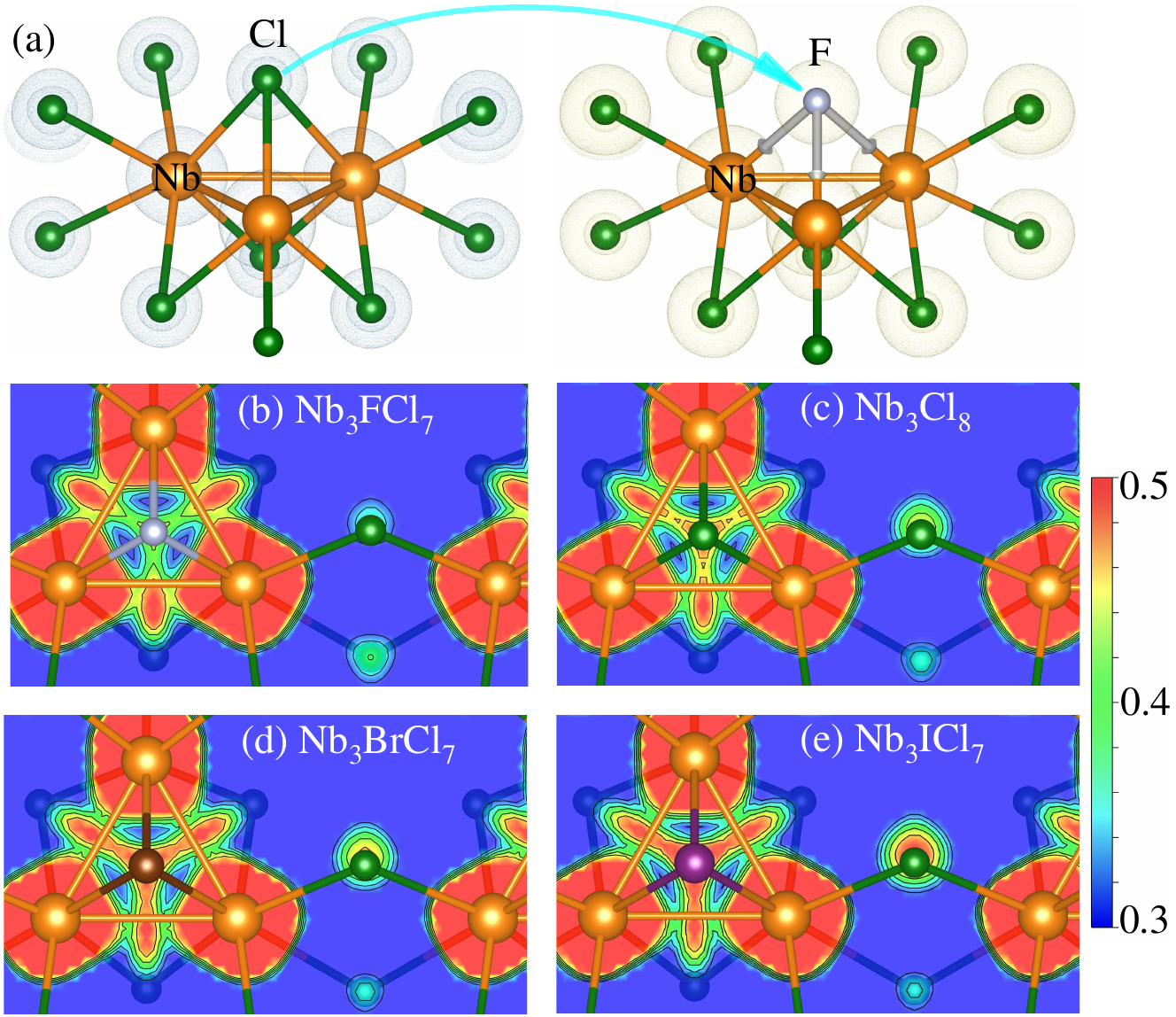]{Fig. 5(a)}). This substitution strategy modifies the bonding radius and orbital size of the apical atoms, thereby tuning the chemical precompression exerted on the Nb trimer. Such precompression subtly modulates the spacing within Nb trimer, ultimately governing the breathing strength. It is worth noting that the concept of chemical precompression was introduced by Ashcroft in his proposal of hydrogen-rich superconductors \cite{PhysRevLett.92.187002}. For instance, at 38 GPa, the average valence electron density $r_s$ of methane ($\rm CH_4$) is equivalent to that of hydrogen under eightfold compression (150 GPa). This indicates that $\rm CH_4$ achieves a precompression effect on hydrogen through C--H chemical bonding (i.e., chemical precompression).

In this paper, based on the first-principles method, we investigate the influences of electron correlation and breathing effect on MIT and magnetic properties of the 2D layered kagome material $\rm Nb_3Cl_8 $. The structure of this paper is organized as follows: In Subsection III-A, we calibrate $U_{\text{eff}}$ by analyzing the electronic structures, optical properties, and magnetic properties, which successfully reconciles the discrepancy between DFT calculation results and experimental measurements in $\rm Nb_3Cl_8 $. In Subsection III-B, we employ Boltzmann transport theory to calculate the electronic transport and magnetic properties of $\rm Nb_3Cl_8 $. In Subsection III-C, we introduce breathing effect into $\rm Nb_3Cl_8 $ by varying the chemical precompression, resulting in $\rm Nb_3XCl_7 $ (X = F, Cl, Br, and I). We calculate the electronic transport and magnetic properties of $\rm Nb_3XCl_7 $. Subsection III-D provides an analysis of the stability and potential synthesis pathways for $\rm Nb_3XCl_7 $. Section IV summarizes our key conclusions: the breathing effect and electron correlation can be utilized to control the band gaps, conductivity, and magnetic properties of kagome materials.

\vspace{-0.2cm}  
\section{II. MODEL AND METHOD}
\vspace{-0.3cm}

First-principles calculations were performed using the Vienna \textit{ab initio} simulation package (\textsc{Vasp})~\cite{PhysRevB.54.11169,CMS.0256} within the projector augmented-wave (PAW) method and the plane-wave basis set, employing the Perdew-Burke-Ernzerhof (PBE) exchange-correlation functional \cite{PhysRevLett.77.3865}. For accurate band structure calculations, the Heyd-Scuseria-Ernzerhof (HSE) screened Coulomb hybrid density functional was employed \cite{10.1063/1.1760074,paier2006screened}. A plane-wave cutoff energy of 400 eV and dense $\mathbf{k}$-point meshes with a spacing of $2\pi \times 0.02 \; \rm{\mathring{A}^{-1}}$ were used to ensure that the total energies of the optimized structures converged within 1~meV/atom. Electron correlation effects are treated using the DFT+$U$ method \cite{PhysRevB.57.1505,PhysRevB.71.035105}, with $U_{\text{eff}}$ applied to the Nb 4$d$ orbitals. To account for the 2D layered structure, a vacuum layer exceeding 25 $\rm \mathring{A}$ was introduced to eliminate spurious interactions between adjacent layers. Structural optimizations were performed with a force convergence criterion of $ 10^{-7}$ $ \rm eV/\mathring{A}$, allowing for full relaxation of both atomic positions and lattice parameters.

The \textsc{BoltzTraP2} software package \cite{MADSEN2018140,murad2025dft} was used to compute the electrical conductivity $\sigma$ and magnetic susceptibility $\chi$. The reliability of this software has been validated by numerous studies \cite{acs.jctc.3c01405,D5RA01965F,acsomega.3c06221}. The \textsc{Vaspkit} code~\cite{VASPKIT} was employed for post-processing analysis. Since deformation potential theory may overestimate relaxation times, the constant relaxation time (CRT) approximation \cite{ZHU2023157817} was adopted. A constant relaxation time of $\tau_0 = 10$ fs at 300 K was used for the kagome monolayers, consistent with values extensively employed in recent studies \cite{Gandi2016,PhysRevB.94.035405,C5NR03813H,ZHU2023157817}.

\vspace{-0.2cm}  
\section{III. RESULT AND DISCUSSION}
\vspace{-0.3cm}

\subsection{A. Addressing the discrepancy between the calculations and experiments}

\begin{figure}[t]
	\centering
	\includegraphics[width=\linewidth]{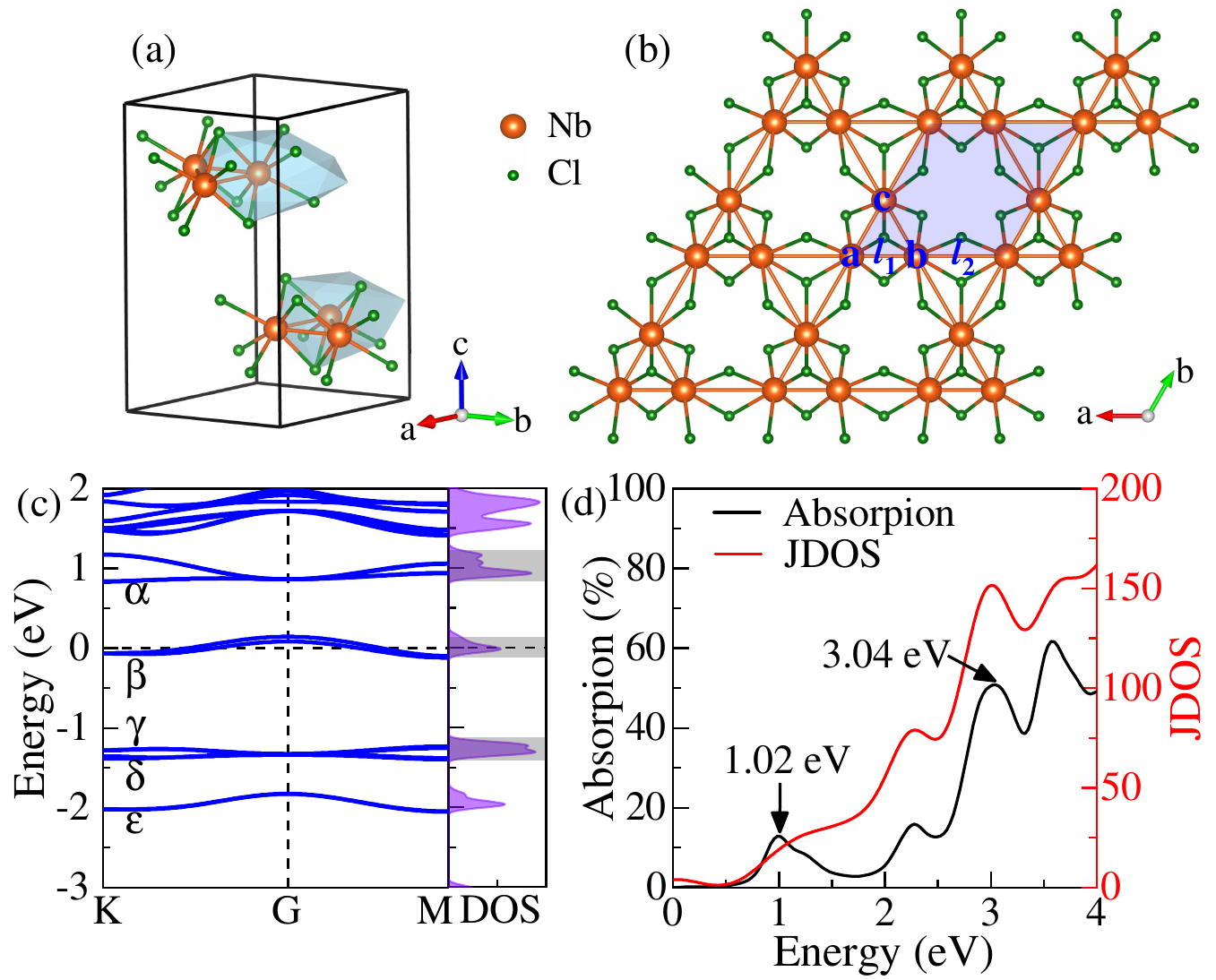}
	\caption{(a) The crystal structure of $\rm Nb_3Cl_8$. The light blue polyhedron represents the coordination polyhedron formed by the central Nb atom and its nearest-neighbor atoms. (b) The top view of the monolayer $\rm Nb_3Cl_8$. The transparent blue region represents a unit cell. The $l_1$ marks the distance between a and b cites in the unit cell. The $l_2$ marks the distance between b cite and the a cite of the next unit cell. (c) The calculated band structures of the monolayer $\rm Nb_3Cl_8$ with the effective correlation strength $U_{\text{eff}}=2$ eV. The light gray regions in the density of states were inferred from the optical absorption spectra \cite{acs.nanolett.2c00778}. (d) The calculated optical absorption and joint density of states of 20 nm $\rm Nb_3Cl_8$ thin film.}
	\label{Fig1.pdf}
\end{figure}

\hyperref[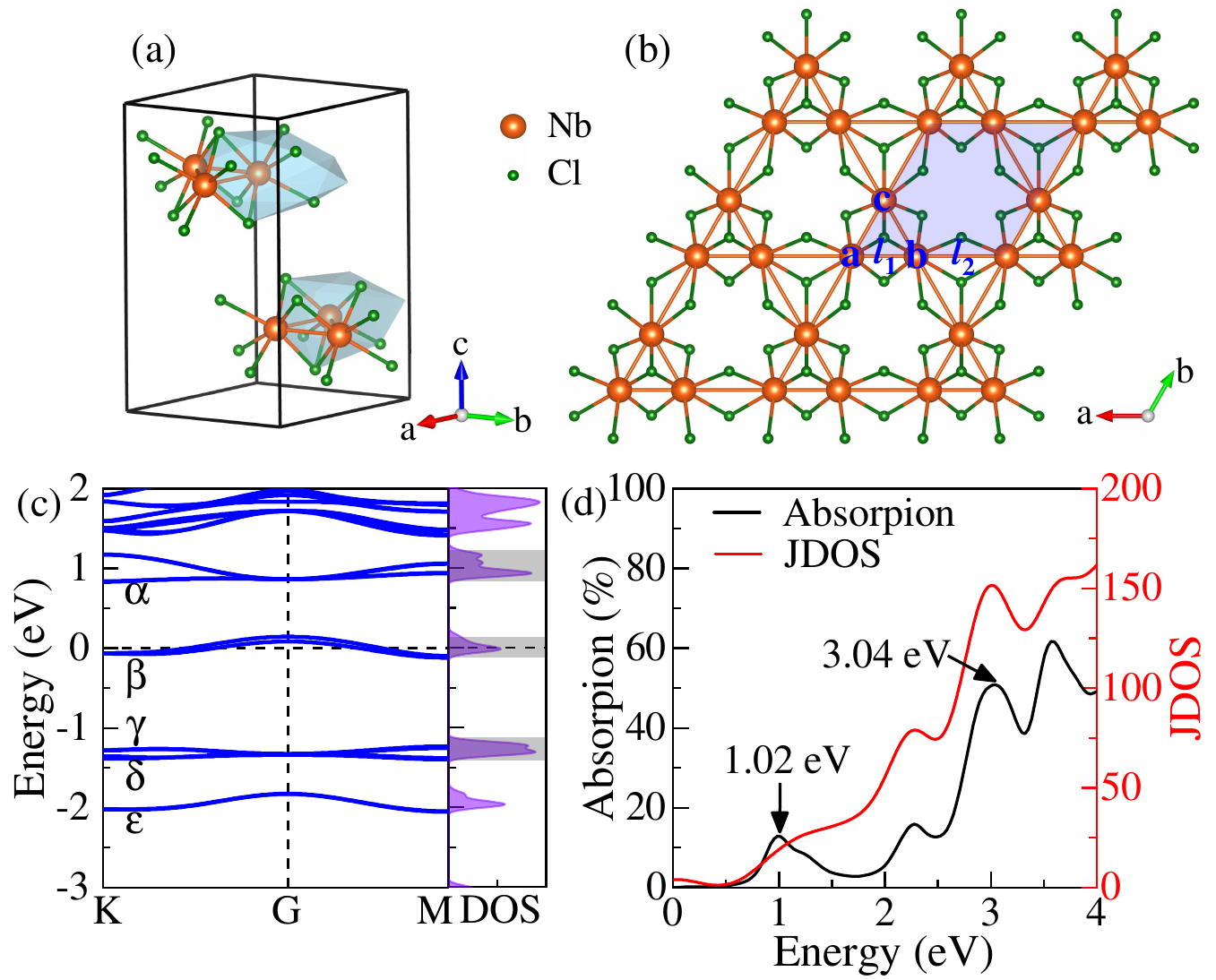]{Fig. 1(a)} presents the structure of $\rm Nb_3Cl_8 $. $\rm Nb_3Cl_8 $ exhibits a typical 2D layered structure, with adjacent layers connected by weak van der Waals interactions along the c-direction. Following structural optimization, $\rm Nb_3Cl_8 $ is determined to crystallize in the $P\overline{3}m1$ space group, with lattice parameters of a = 6.827 $\rm \mathring{A}$ and c = 13.530 $\rm \mathring{A}$, which are close to the experimental values \cite{acs.nanolett.2c00778,Jeff_2023}. $\rm Nb_3Cl_8 $ thin flakes can be mechanically exfoliated onto various substrates, including $\rm SiO_2/Si$, $\rm Au/SiO_2/Si$ \cite{acs.nanolett.2c00778}. The structure of monolayer $\rm Nb_3Cl_8 $ is shown in \hyperref[Fig1.pdf]{Fig. 1(b)}. Three $\rm Nb^{3+}$ ions (labeled a, b, and c) form Nb trimers through lattice distortion. This leads to the distance $l_1$ is smaller than the distance $l_2$ , thereby forming a breathing kagome lattice. Due to the breathing anisotropy, the ratio $l_1/l_2$ = 0.7087, leading to alternating hopping strengths between neighboring sites. The Nb trimers are weakly bonded to Cl atoms (As shown in \hyperref[Fig5.pdf]{Fig. 5(c)}). By constructing a vacuum layer of 25 $\rm \mathring{A}$, we obtained the structure of monolayer $\rm Nb_3Cl_8 $. The calculated phonon dispersion curves for monolayer $\rm Nb_3Cl_8 $ are presented in \hyperref[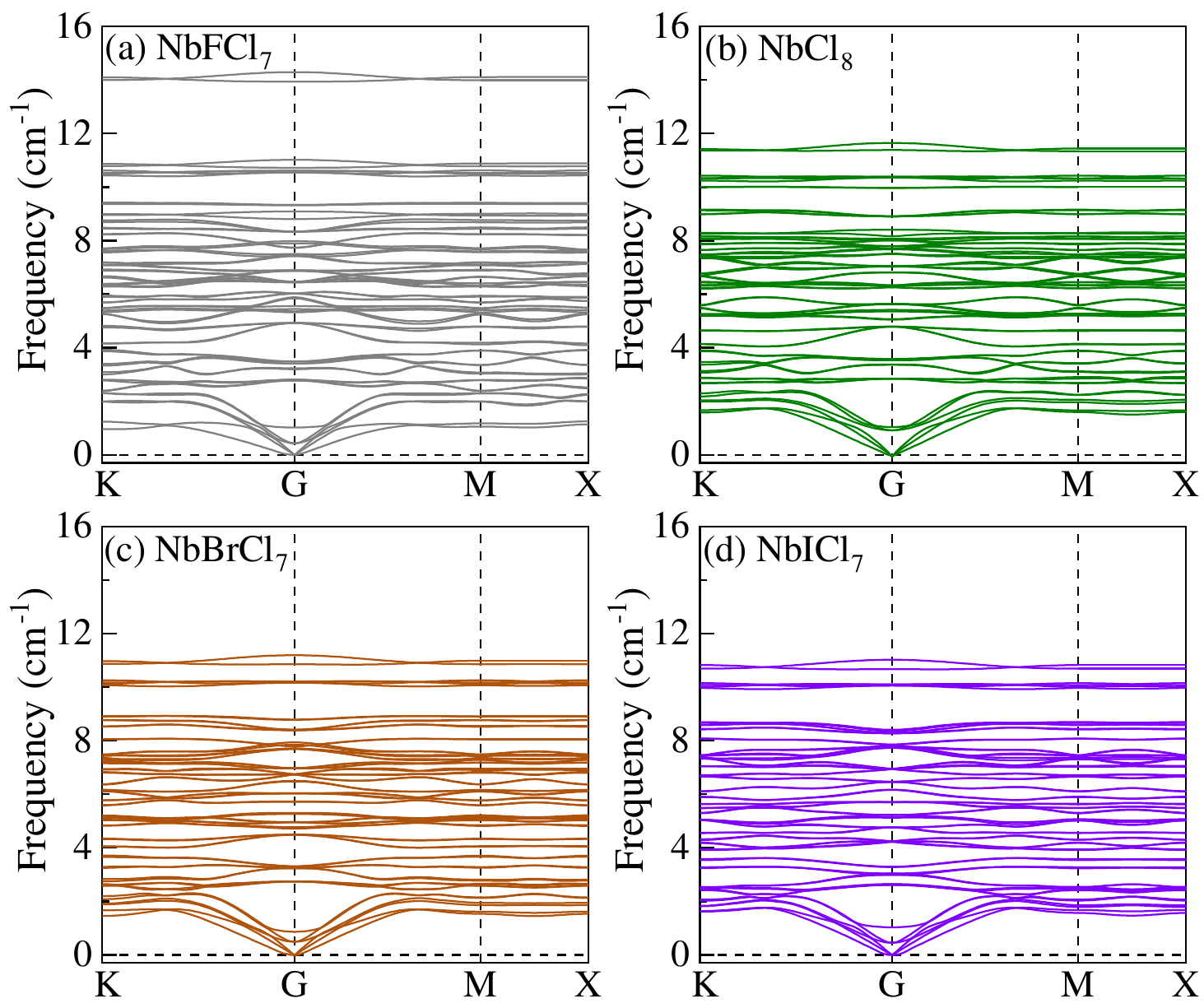]{Fig. 4(b)}. The absence of imaginary frequencies confirms that the system is dynamically stable.

Before delving into the electronic properties of $\rm Nb_3Cl_8 $, it is crucial to address a significant discrepancy observed near the Fermi level: the band gap calculated by DFT is smaller than that measured by optical absorption spectroscopy \cite{acs.nanolett.2c00778}. Specifically, the optical band gap is approximately 1.12 eV, which is about 0.23 eV larger than the calculated value. This discrepancy may arise from the fact that Nb atoms possess fully occupied 3\textit{d} orbitals and nearly half-filled 4\textit{d} orbitals. The presence of numerous localized \textit{d} orbitals could lead to electron correlation effect \cite{PhysRevB.96.195102,PhysRevB.107.125124}, thereby influencing the calculated band structure.

\begin{figure}[t]
	\centering
	\includegraphics[width=\linewidth]{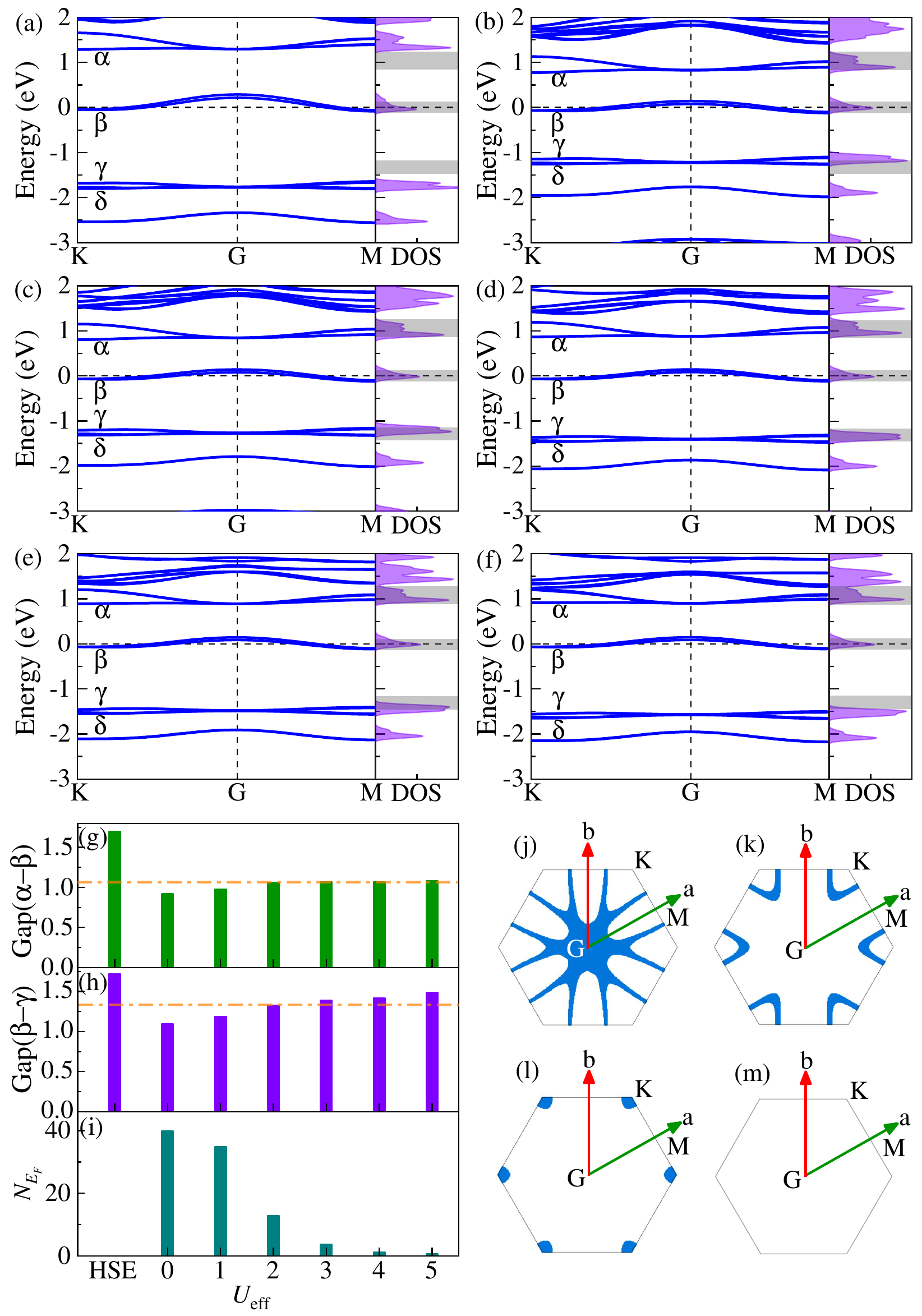}
	\caption{(a) The calculated electronic structures of monolayer $\rm Nb_3Cl_8 $ with the Heyd-Scuseria-Ernzerhof hybrid functional. The light gray regions in the density of states were inferred from the optical absorption spectra \cite{acs.nanolett.2c00778}. (b)-(f) The calculated electronic structures of monolayer $\rm Nb_3Cl_8 $ with the effective correlation strength $U_{\text{eff}}=0,1,3,4,5$ eV. The light gray regions in density of states were inferred from the optical absorption spectra \cite{acs.nanolett.2c00778}. (g)-(h) The calculated band gaps ($\rm \alpha-\beta$) and band gaps ($\rm \beta-\gamma$). The orange dashed lines represent the experimentally measured band positions \cite{acs.nanolett.2c00778}. (i) The calculated density of states at the Fermi level of monolayer $\rm Nb_3Cl_8 $. (j)-(m) The Fermi lines of monolayer $\rm Nb_3Cl_8 $ with $U_{\text{eff}}=0,1,2,3$ eV.}
	\label{Fig2.pdf}
\end{figure}

In an attempt to obtain a calculated band structure that aligns with experimental results, we employed the HSE hybrid functional and the DFT+$U$ method separately to calculate the band structure of $\rm Nb_3Cl_8 $. The electronic structure obtained by the HSE hybrid functional is shown in \hyperref[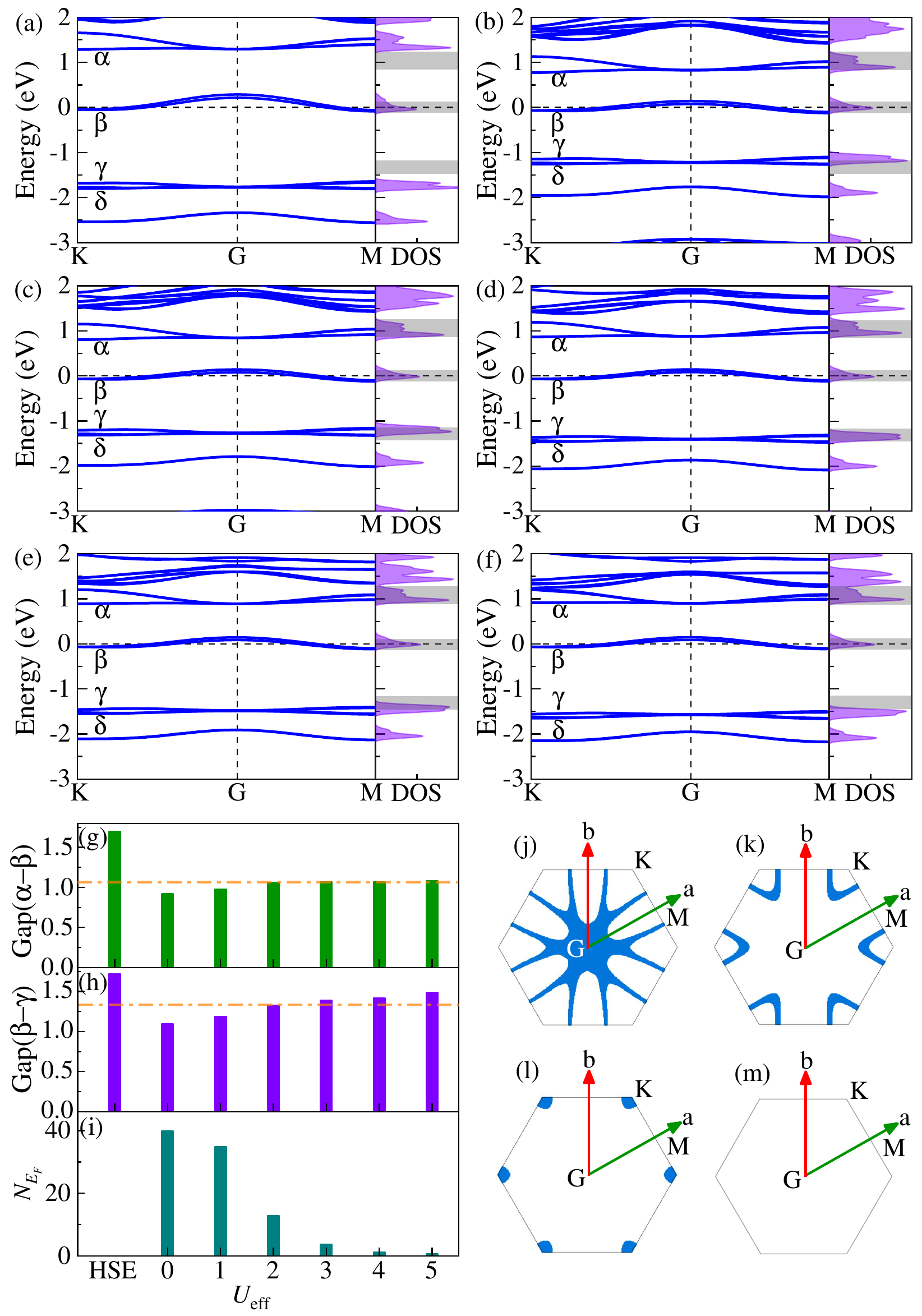]{Fig. 2(a)}. The $\rm \delta$ band corresponds to the characteristic flat band of the kagome lattice, while the combination of the $\rm \gamma$ and $\rm \beta$ bands forms the Dirac cone opened by the breathing effect. Our previous work \cite{cpl_42_9_090712} demonstrated the the Dirac cone opened by the breathing effect within the tight-binding model of the kagome lattice. The light gray regions in the density of states (DOS) were inferred from the optical absorption spectra \cite{acs.nanolett.2c00778}. It is evident that the $\rm \alpha$, $\rm \gamma$ and $\rm \delta$ bands deviate significantly from the experimental results, suggesting that the HSE hybrid functional may not be suitable for this system. The calculated band structures with $U_{\text{eff}}$ values of 0, 1, 2, 3, 4, and 5 eV are presented in \hyperref[Fig2.pdf]{Figs. 2(b)-(c)}, \hyperref[Fig1.pdf]{Fig. 1(c)}, and \hyperref[Fig2.pdf]{Figs. 2(d)-(f)}, respectively. We summarize the Gap($\rm \alpha-\beta$) and Gap($\rm \beta-\gamma$) in \hyperref[Fig2.pdf]{Figs. 2(g)-(h)}. The orange dashed lines represent the experimentally measured band gaps \cite{acs.nanolett.2c00778}. The results indicate that both Gap($\rm \alpha-\beta$) and Gap($\rm \beta-\gamma$) are underestimated when $U_{\text{eff}}$ = 0 and 1 eV. However, the band gaps align well with the experimental data when $U_{\text{eff}}= 2$ eV. Upon further increasing the correlation strength, Gap($\rm \beta-\gamma$) gradually becomes overestimated. Therefore, the electronic structure of the $\rm Nb_3Cl_8 $ system is most consistent with experimental results when the $U_{\text{eff}}= 2$ eV.

To further validate the choice of $ U_{\text{eff}} = 2 $ eV for $\rm Nb_3Cl_8$, we have additionally calculated its optical absorption, joint density of states (JDOS) and magnetic ground state. \hyperref[Fig1.pdf]{Fig. 1(d)} displays the optical absorption of  $\rm Nb_3Cl_8$ thin film. We adopted a thickness $ d = 20 $ nm consistent with the experimental value \cite{acs.nanolett.2c00778}. A prominent absorption peak is observed near 1.02 eV, with an absorption rate of $ 15\% $. These features are in good agreement with experimental results. It is worth noting that the film exhibits an absorption rate of $ 51\% $ near 3.04 eV in the violet light region. The remarkably high absorption rate is attributed to the strong electron localization of the flat band $\delta$ (As shown in \hyperref[Fig1.pdf]{Fig. 1(c)}), which provides a high DOS accessible to violet light. We further calculated the JDOS of monolayer $\rm Nb_3Cl_8$ and display it in \hyperref[Fig1.pdf]{Fig. 1(d)}. A rapid increase in JDOS is observed near 1.05 eV, and a broad peak appears near 3.04 eV. These results are strongly correlated with the optical absorption and are fully consistent with Fermi's golden rule \cite{adma.201306281}. The aforementioned results indicate that $\rm Nb_3Cl_8$ film has the potential to become the novel optoelectronic devices.

Furthermore, we have performed complete structural optimizations for the three distinct magnetic configurations: FM, AFM, and NM. The corresponding energies are denoted as $ E_{\rm FM}=-109.204 $ eV, $ E_{\rm AFM}=-109.205 $ eV, and $ E_{\rm NM}=-108.960 $ eV. The lowest energy of the AFM state indicates that the system exhibits antiferromagnetism at low temperatures. Experimentally, by fitting the Curie-Weiss law to $ \rm Nb_3Cl_8$ \cite{haraguchi2017}, the researchers obtained a Curie-Weiss temperature of $\Theta_{\text{W}} = -13.1(3)$ K. The negative $ \Theta_{\text{W}} $ indicates that the magnetic interactions between Nb trimers are predominantly AFM. Thus, the calculated magnetic ground state is in agreement with the experimental results. In summary, based on the electronic structure, optical absorption, and magnetic ground state, we ultimately determined $ U_{\text{eff}} $ to be 2 eV.

\vspace{-0.2cm}
\subsection{B. The regulation of electronic correlation}
\vspace{-0.2cm}

Previous studies using the DQMC method \cite{cpl_42_9_090712} indicated that the interplay between the breathing effect and electronic correlation drives the kagome system to transition from a PM metal to a Mott insulator. Building upon this foundation, we investigate the regulatory roles of electron correlation and breathing effect in the synthesized material $\rm Nb_3Cl_8$. Initially, we focus on the impact of electron correlation on the monolayer $\rm Nb_3Cl_8$. 

\begin{figure}[t]
	\centering
	\includegraphics[width=\linewidth]{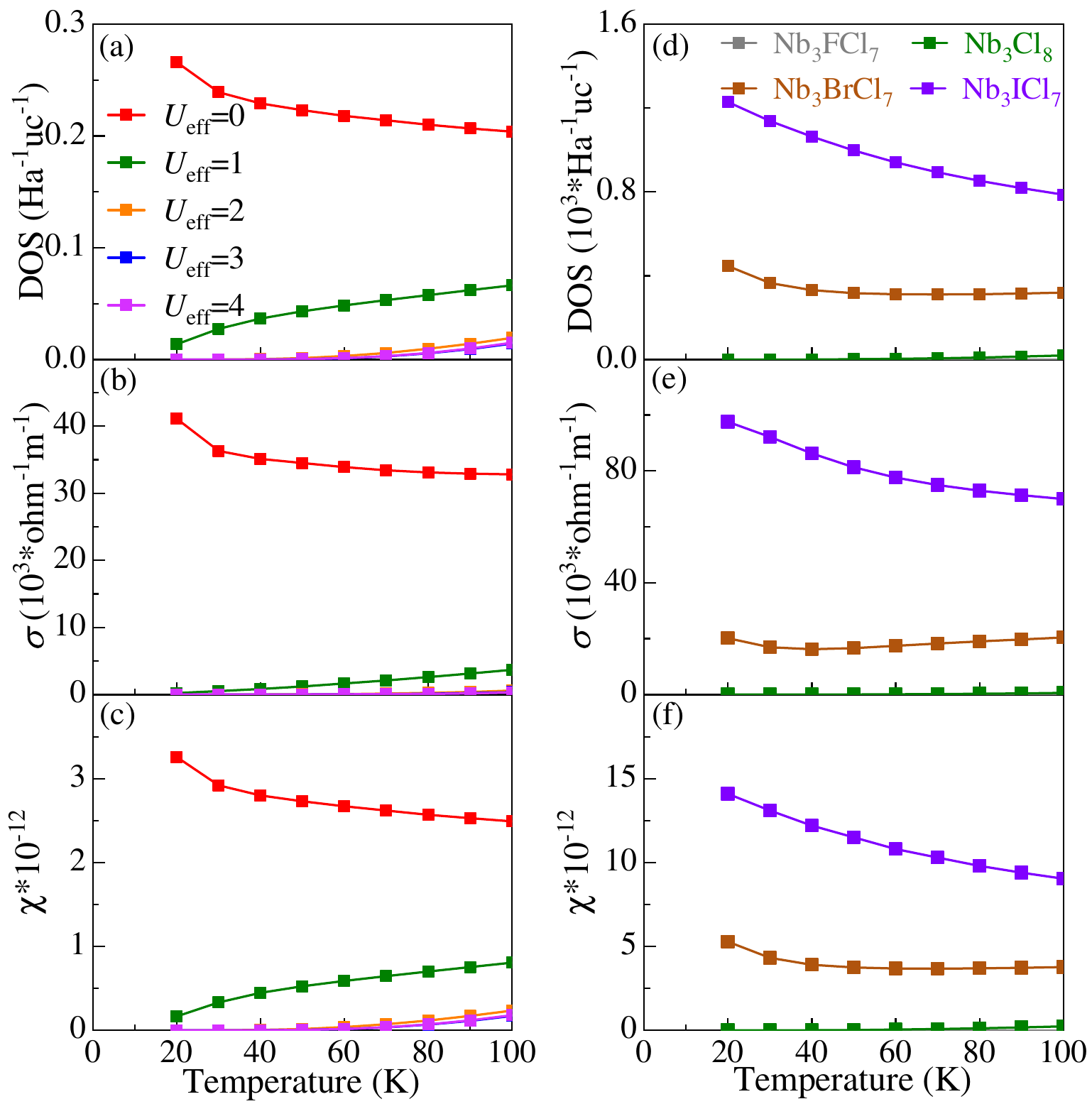}
	\caption{(a)-(c) The density of states, conductivity, and magnetic susceptibility of monolayer $\rm Nb_3Cl_8 $ with the effective correlation strength $U_{\text{eff}}=0,1,3,4$ eV and the temperature range from 20 K to 100 K. (d)-(f) The density of states, conductivity, and magnetic susceptibility of the monolayer $\rm Nb_3XCl_7$ (X = F, Cl, Br, and I) with the temperature range from 20 K to 100 K.}
	\label{Fig3.pdf}
\end{figure}

We employ the Boltzmann transport theory to calculate the electronic transport and magnetic properties of the $\rm Nb_3Cl_8$ with electron correlation. We begin with structural optimization, setting $U_{\text{eff}}$ to 0, 1, 2, 3, and 4 eV to achieve a systematic variation of electron correlation. Given the known discrepancy between experimental and theoretical Fermi levels \cite{acs.nanolett.2c00778}, we uniformly shift the chemical potential upward by 0.88 eV.  Since the temperature dependence of $\sigma$ at low temperatures can reflect the MIT of the system, we set the temperature range from 20 K to 100 K. The DOS, $\sigma $ and $\chi $ are presented in \hyperref[Fig3.pdf]{Figs. 3(a)-(c)}. The $\sigma $ is obtained by multiplying the calculated $\sigma /\tau_0$ by $\tau_0$. Moreover, we present the DOS at the Fermi level $N_{E_F}$ and the Fermi lines at 0 K in \hyperref[Fig2.pdf]{Fig. 2(i)} and \hyperref[Fig2.pdf]{Figs. 2(j)-(m)}, respectively. It can be observed that as $ U_{\text{eff}} $ increases, $N_{E_F}$ decreases rapidly, and the hole-like Fermi lines shrinks quickly along the $\Gamma$-K direction. Those results suggest that the $\sigma $ of $\rm Nb_3Cl_8$ at 0 K may decrease sharply with increasing $ U_{\text{eff}} $. \hyperref[Fig3.pdf]{Fig. 3(a)} indicates that as the temperature rises, the DOS with $ U_{\text{eff}} = 0 $ eV gradually decreases, whereas the DOS exhibits an upward trend for higher $ U_{\text{eff}} $. Furthermore, the $\sigma $ gradually decreases with increasing temperature for $U_{\text{eff}} = 0$ eV (As shown in \hyperref[Fig3.pdf]{Fig. 3(b)}), indicating metallic behavior. For $ U_{\text{eff}} = 1,2,3,4 $ eV, the $\sigma $ gradually increase with increasing temperature and are much smaller than that for $U_{\text{eff}} = 0$ eV. This indicates that the system transitions to an insulator. The results for both DOS and $\sigma $ indicate that the monolayer $\rm Nb_3Cl_8$ is metallic at $ U_{\text{eff}} = 0 $ eV, whereas it transitions to an insulator at $ U_{\text{eff}}=1 $ eV.

On the other hand, \hyperref[Fig3.pdf]{Fig. 3(c)} presents the $\chi$ of monolayer $\rm Nb_3Cl_8 $ for various $ U_{\text{eff}} $ values. For $ U_{\text{eff}} = 0 $ eV, the magnitude of $\chi$ is on the order of $10^{-12}$, and it decreases slowly with increasing temperature. This indicates that the system may exhibit AFM behavior in the ground state. However, above 20 K, the temperature exceeds the Néel temperature, and the system exhibits PM characteristics. For $ U_{\text{eff}} > 0 $ eV, $\chi$ remain small in magnitude and increase with increasing temperature, which is characteristic of antiferromagnets. Overall, monolayer $\rm Nb_3Cl_8$ undergoes a transition from a PM metal to a Mott insulator at $ U_{\text{eff}} = 1 $ eV. This is consistent with the results predicted by DQMC method \cite{cpl_42_9_090712}.

\begin{figure}[t]
	\centering
	\includegraphics[width=\linewidth]{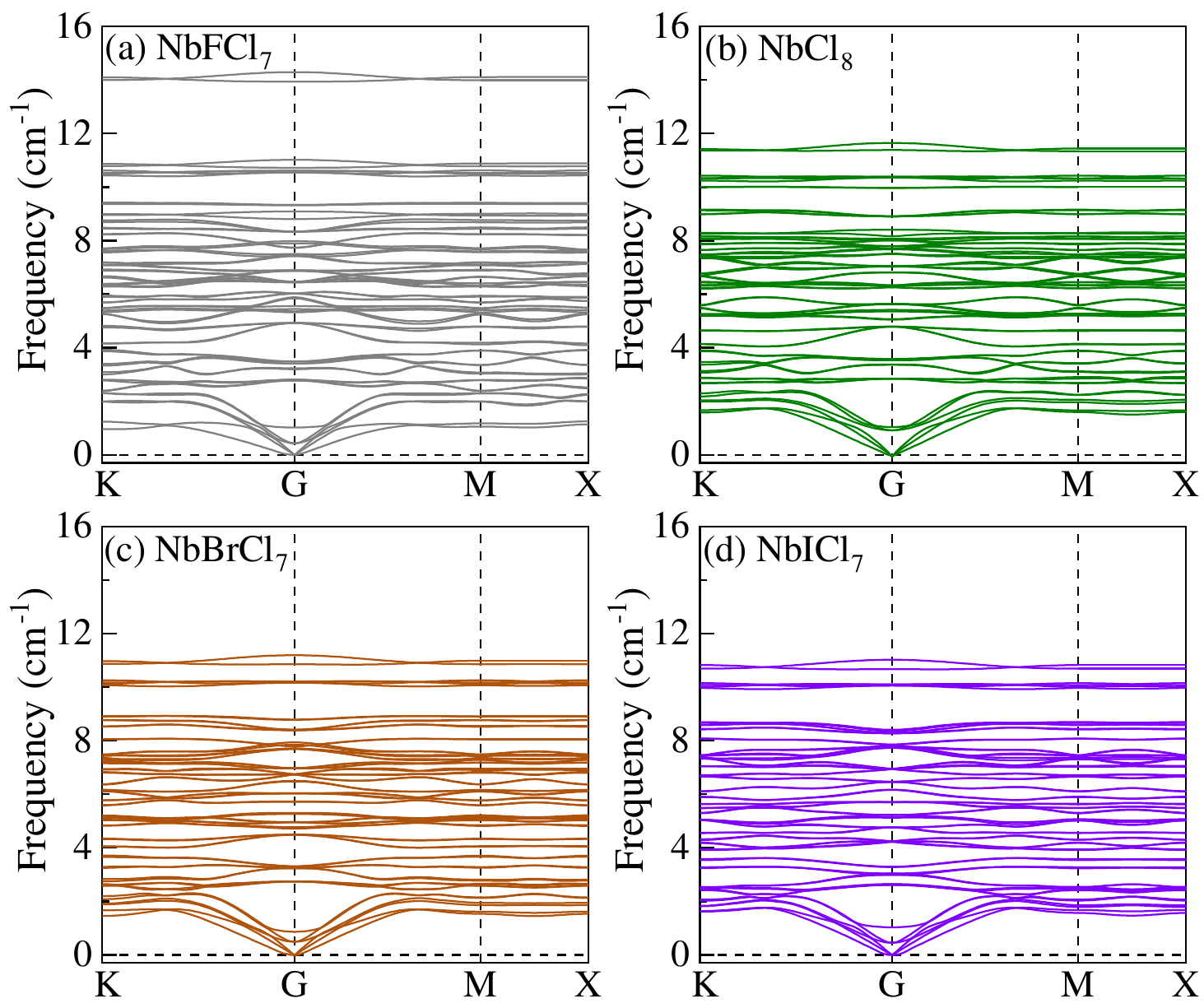}
	\caption{The phonon dispersion curves for the monolayer $\rm Nb_3XCl_7$ (X = F, Cl, Br, and I).}
	\label{Fig4.pdf}
\end{figure}

\subsection{C. The regulation of breathing interaction}

\begin{figure}[t]
	\centering
	\includegraphics[width=\linewidth]{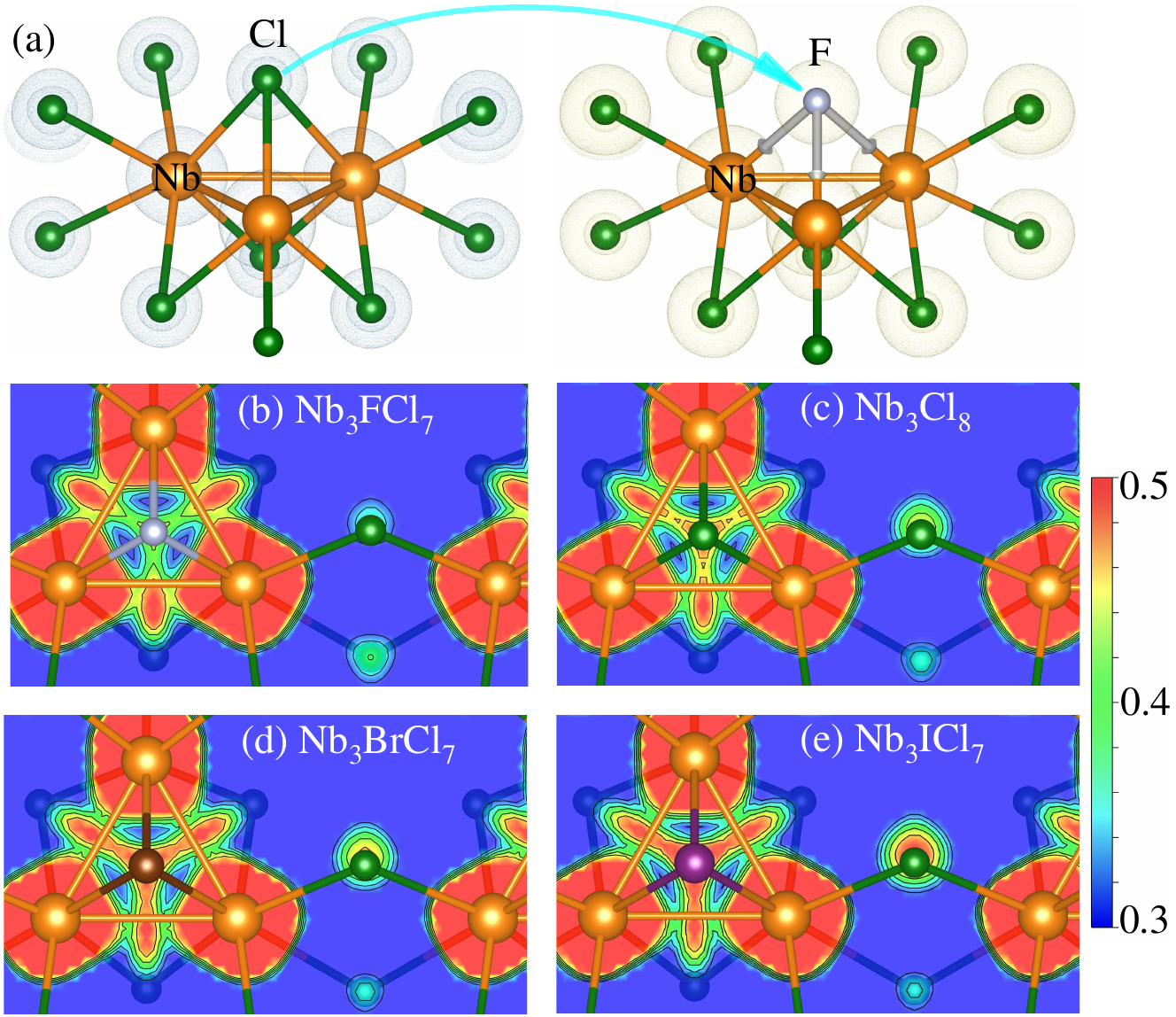}
	\caption{(a) The schematic representation of the crystal structure transition from $\rm Nb_3Cl_8$ to $\rm Nb_3FCl_7$. (b)-(d) The contour maps of the electron localization functions on the Nb trimer plane for $\rm Nb_3XCl_7$ (X = F, Cl, Br, and I).}
	\label{Fig5.pdf}
\end{figure}

We now investigate the impact of breathing effect on the key properties of $\rm Nb_3Cl_8$ under reasonable electronic correlation strengths ($U_{\text{eff}} = 2$ eV). We take $\rm Nb_3Cl_8$ as the base structure and replace the Cl atoms at the top of the Nb trimer with F, Br, and I. After structural optimization, we obtained a series of breathing kagome systems: $\rm Nb_3XCl_7$ (X = F, Cl, Br, and I). The phonon dispersion curves for these structures are presented in \hyperref[Fig4.pdf]{Fig. 4}. The absence of imaginary frequencies indicates that all structures are dynamically stable. In Subsection III-D, we will further discuss the feasibility of experimentally synthesizing $\rm Nb_3XCl_7$. Taking the transition from $\rm Nb_3Cl_8$ to $\rm Nb_3FCl_7$ as an example, we illustrate this process in the crystal diagram (As shown in \hyperref[Fig5.pdf]{Fig. 5(a)}). By altering the bonding radius of the top atoms (from Cl atom to F atom), we aim to change the chemical precompression exerted by the top atoms on the Nb trimers. This slightly alter the atomic spacing within the Nb trimer, thereby regulating the intensity of the breathing effect. Typically, the breathing strength is characterized by the ratio $l_1/l_2$ \cite{PhysRevLett.118.237203,PhysRevB.104.L060405}. \hyperref[Table1]{Tab. 1} presents the $l_1/l_2$ ratios and the corresponding covalent radii $r$ of the X elements in $\rm Nb_3XCl_7$ \cite{B801115J}. As $r$ increases from F to I, the $l_1/l_2$ ratios $\rm Nb_3XCl_7$ also increase. This is because a larger covalent radius weakens the attraction of the atomic nucleus to valence electrons, which in turn reduces the bonding strength between the top atoms and the Nb trimers. Consequently, this leads to a weaker confinement of the Nb trimers, i.e., an increase in $l_1/l_2$.

Given that the breathing strength in the DQMC method is determined by the ratio of the intracell hopping strength $t_{\text{in}}$ to the intercell hopping strength $t_{\text{out}}$ \cite{cpl_42_9_090712}, we attempt to establish a breathing strength analogous to $t_{\text{out}}/t_{\text{in}}$. The electron localization function (ELF) is defined as \cite{10.1063/1.458517,Silvi1994}:

\begin{align}
\text{ELF} & = (1 + \chi_{\sigma}^2)^{-1} \\
\chi_{\sigma} & = D_\sigma/D_\sigma^0 \\
D_\sigma & = \sum_i |\nabla \varphi_i|^2-\frac{(\nabla \rho_\sigma)^2}{4\rho_\sigma} \\
D_\sigma^0 & = \frac{3}{5}(6\pi^2)^{2/3}\rho_\sigma^{5/3}
\end{align}

\noindent Here, $\varphi_i$ denotes the single-electron wavefunction and $\rho_\sigma$ is the $\sigma$-spin density. When $D_\sigma=D_\sigma^0$, the electron state in this region resembles a uniform electron gas, which easily conduct electrons. In this case, $\text{ELF}=0.5$. Therefore, we postulate that the hopping strength $t$ between lattice sites is the integral of ELF $\in (0.3, 0.5)$ between lattice sites:

\begin{equation}
t \propto \iint_{\text{site1}}^{\text{site2}} \text{ELF}  \, dS
	\label{Eq5}
\end{equation}

\noindent \hyperref[Fig5.pdf]{Figs. 5(b)-(d)} presents the contour maps of the ELF on the Nb trimer plane. Utilizing this ELF contour map, we evaluated \hyperref[Eq5]{Eq. 5} to obtain the $t$ values. Furthermore, we calculated $t_{\text{in}}$ and $t_{\text{out}}$, and their ratio for $\rm Nb_3XCl_7$ are shown in \hyperref[Table1]{Tab. 1}. The results show that the trend of $t_{\text{out}}/t_{\text{in}}$ with varying of the top atoms is consistent with that of $l_1/l_2$. Therefore, in the following articles, we will use $t_{\text{out}}/t_{\text{in}}$ to characterize the breathing strength.

\begin{table}[h]
	\centering
	\caption{Calculated $l_1/l_2$ ratios, covalent radii $r$, and $t_{\text{out}}/t_{\text{in}}$ ratios for Nb$_3$XCl$_7$ systems.}
	\begin{tabular}{|c|c|c|c|}
	\hline
	X & $l_1 / l_2$ & $r$ ($\rm \mathring{A}$) & $t_{\text{out}}/t_{\text{in}}$ \\
	\hline
	F & 0.7009 & 0.72 & 0.2759 \\
	Cl & 0.7087 & 0.99 & 0.5891 \\
	Br & 0.7102 & 1.21 & 0.6674 \\
	I & 0.7103 & 1.48 & 0.7527 \\
	\hline
	\end{tabular}
	\label{Table1}
\end{table}

Next, \hyperref[Fig3.pdf]{Figs. 3(d)-(f)} demonstrate the electronic transport and magnetic properties of $\rm Nb_3XCl_7$ with breathing effect. \hyperref[Fig3.pdf]{Fig. 3(e)} shows that the average conductivity of $\rm Nb_3ICl_7$ is $8 \times 10^4 \, \Omega^{-1} \text{m}^{-1}$ and $\sigma$ gradually decreases with increasing temperature. This indicates that $\rm Nb_3ICl_7$ with $t_{\text{out}}/t_{\text{in}} = 0.7527$ exhibits metallic behavior. When $t_{\text{out}}/t_{\text{in}}$ decreases to 0.6674 for $\rm Nb_3BrCl_7$, the average conductivity decreases to $2 \times 10^4 \, \Omega^{-1} \text{m}^{-1}$ and $\sigma$ shows little temperature dependence. This suggests that $\rm Nb_3BrCl_7$ might be at the boundary between a metal and an insulator. As $t_{\text{out}}/t_{\text{in}}$ further decreases to 0.5891 and 0.2759 for $\rm Nb_3Cl_8$ and $\rm Nb_3FCl_7$, respectively, their $\sigma$ both drop to nearly zero and the systems transition to insulators. These results are also supported by the DOS at the Fermi level (As shown in \hyperref[Fig3.pdf]{Fig. 3(d)}). The average DOS for $\rm Nb_3ICl_7$ and $\rm Nb_3BrCl_7$ reaches 900 and 400 states/Ha/uc, respectively. However, the average DOS for $\rm Nb_3Cl_8$ and $\rm Nb_3FCl_7$ nearly vanishes. This indicates that as the breathing effect intensifies, the DOS will continuously decrease to zero, driving $\rm Nb_3XCl_7$ to transition to an insulator. Therefore, $\rm Nb_3BrCl_7$ with $t_{\text{out}}/t_{\text{in}}=0.6674$ might represent a critical point for a bandwidth-controlled MIT. Theoretical research using the DQMC method has shown that the enhancement of the breathing effect can gradually transform the kagome system into an insulator \cite{cpl_42_9_090712}. The study on breathing kagome systems $\rm Nb_3XCl_7$ provides strong verification of this conclusion.

Subsequently, we investigate the impact of breathing effect on magnetic properties. \hyperref[Fig3.pdf]{Fig. 3(f)} shows the $\chi$ for $\rm Nb_3XCl_7$ under four different breathing strength. The $\chi$ of $\rm Nb_3ICl_7$ and $\rm Nb_3BrCl_7$ decreases slowly with increasing temperature. This indicates that $\rm Nb_3ICl_7$ and $\rm Nb_3BrCl_7$ may exhibit PM behavior above 20 K. On the other hands, the $\chi$ of $\rm Nb_3Cl_8$ and $\rm Nb_3FCl_7$ are close to zero, suggesting AFM characteristics. To sum up, $\rm Nb_3XCl_7$ undergoes a transition from PM to AFM behavior near $t_{\text{out}}/t_{\text{in}} = 0.6674$. These results are consistent with the theoretical research using DQMC method \cite{cpl_42_9_090712}.

\begin{figure}[t]
	\centering
	\includegraphics[width=\linewidth]{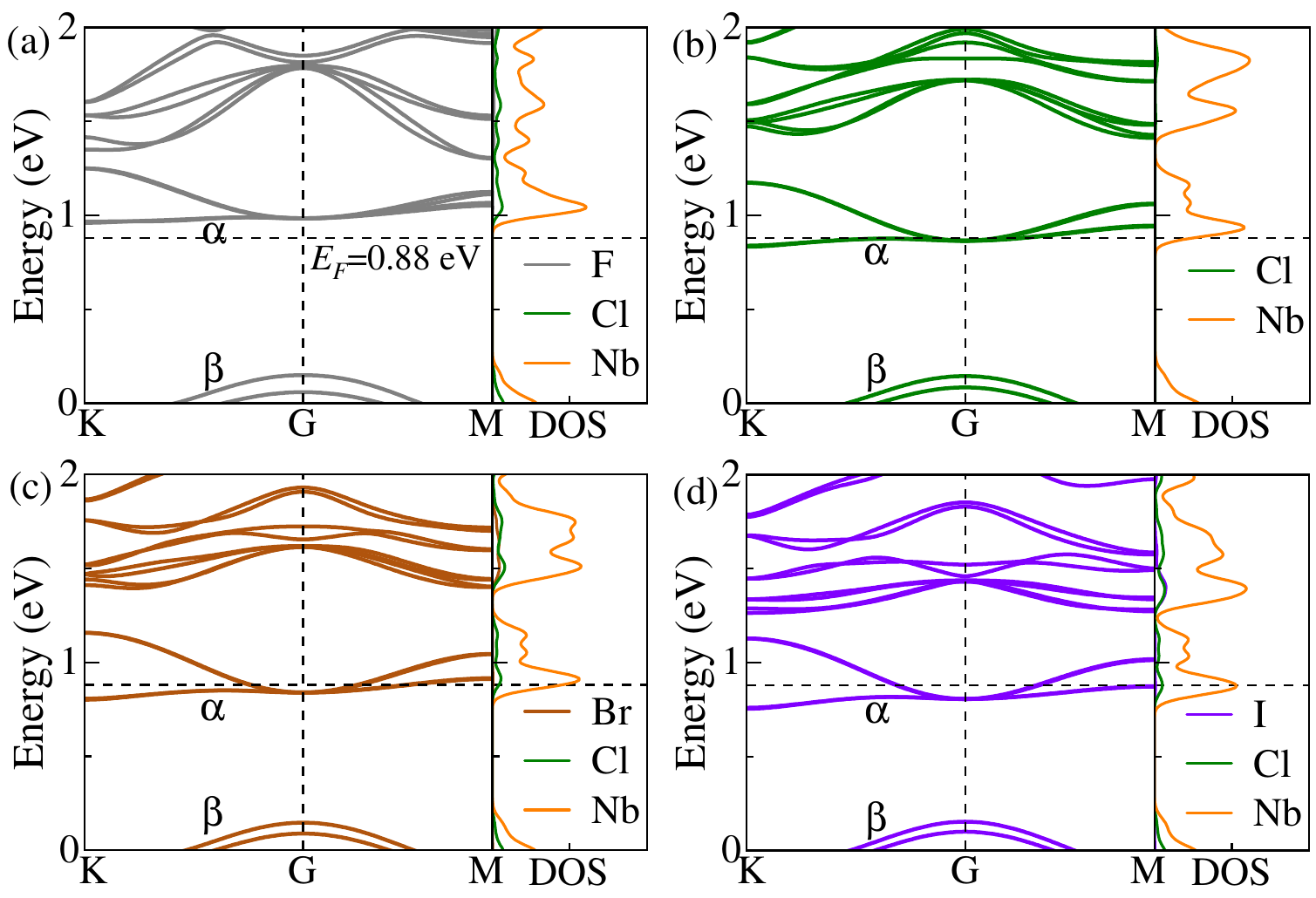}
	\caption{The calculated band structures and projected electronic density of states for $\rm Nb_3XCl_7$ (X = F, Cl, Br, and I). Given the known discrepancy between experimental and theoretical Fermi levels \cite{acs.nanolett.2c00778}, the Fermi levels are shifted upward by 0.88 eV and marked by the dashed line.}
	\label{Fig6.pdf}
\end{figure}

To investigate the mechanism by which the breathing effect controls the MIT in $\rm Nb_3XCl_7 $ systems, we calculated their band structures and projected electronic density of states (PDOS) and show them in \hyperref[Fig6.pdf]{Fig .6}. Given the known discrepancy between experimental and theoretical Fermi levels \cite{acs.nanolett.2c00778}, the Fermi levels are shifted upward by 0.88 eV. \hyperref[Fig6.pdf]{Fig. 6(d)} indicates that for $\rm Nb_3ICl_7$ with $t_{\text{out}}/t_{\text{in}}=0.7527$, the Fermi level lies at the center of the conduction band and coincides with the peak of the PDOS for Nb. The high DOS at the Fermi level implies that $\rm Nb_3ICl_7 $ possesses a sufficient number of conduction electrons, thereby exhibiting high conductivity.
As $t_{\text{out}}/t_{\text{in}}$ decreases, the conduction bands and the PDOS for Nb in $\rm Nb_3BrCl_7$, $\rm Nb_3Cl_8$, and $\rm Nb_3FCl_7$ shift upward, moving away from the Fermi level (As shown in \hyperref[Fig6.pdf]{Figs. 6(a)-(c)}). This causes $\rm Nb_3XCl_7 $ to gradually lose conduction electrons, thereby transitioning to insulator. Furthermore, the ELF in \hyperref[Fig5.pdf]{Figs. 5(b)-(e)} provide direct visualization of how the breathing effect promotes the MIT in $\rm Nb_3XCl_7 $ systems. The results indicate that $\rm Nb_3ICl_7 $, with the weakest breathing strength, possesses the most extensive region of free electron gas between intercell Nb atoms. This arises from the weaker chemical precompression exerted by I atoms, which maximizes the distance $l_2$ between intercell Nb atoms, thereby facilitating the formation of a free electron gas. As $t_{\text{out}}/t_{\text{in}}$ decrease, $l_2$ increases and the region of free electron gas diminishes. This gradually eliminates the conductivity of $\rm Nb_3XCl_7$. The above study indicates that by exploiting the characteristics of atoms in the same group having the same valence but different covalent radii and bonding strengths, one can finely adjust the breathing strength, thereby controlling the MIT in the system.

\subsection{D. Potential synthetic routes}

Finally, we assess the experimental feasibility of synthesizing $\rm Nb_3XCl_7 $ (X = F, Cl, Br, I). Layered $\rm Nb_3Cl_8$ has been synthesized via solid-state reaction \cite{acs.nanolett.2c00778}. Specifically, high-purity Nb and $\rm NbCl_5$ are mixed, heated, and then cooled continuously, eventually using blue Nitto tapes to obtain thin flakes of $\rm Nb_3Cl_8$. $\rm Nb_3TeCl_7$ has been prepared using a similar approach \cite{adma.202301790}, involving the mixture of high-purity Nb, tellurium (Te), and $\rm NbCl_5$, followed by heating and cooling. Drawing upon these established synthetic protocols, we designed several potential pathways for synthesizing $\rm Nb_3FCl_7$, $\rm Nb_3BrCl_7$, and $\rm Nb_3ICl_7$, and calculated their enthalpies of formation. For the gas-solid reaction method, only the pathway to $\rm Nb_3FCl_7$ is exothermic, whereas those to $\rm Nb_3BrCl_7$ and $\rm Nb_3ICl_7$ are endothermic, as detailed below:

\begin{equation}
\frac{8}{5} \text{Nb} + \frac{7}{5} \text{NbCl}_5 + \frac{1}{2} \text{F}_2 \rightarrow \text{Nb}_3\text{FCl}_7 + 0.230 \, \text{eV}
\end{equation}

For the gas-solid phase displacement reaction method, the pathways to $\rm Nb_3FCl_7$ and $\rm Nb_3BrCl_7$ are both exothermic, as detailed below:

\begin{equation}
\text{Nb}_3\text{Cl}_8 + \frac{1}{2} \text{F}_2 \rightarrow \text{Nb}_3\text{FCl}_7 + \frac{1}{2} \text{Cl}_2\uparrow + 2.078 \, \text{eV}
\end{equation}

\begin{equation}
\text{Nb}_3\text{Cl}_8 + \text{Br} \rightarrow \text{Nb}_3\text{BrCl}_7 + \frac{1}{2} \text{Cl}_2\uparrow + 0.168 \, \text{eV}
\end{equation}

Since no exothermic pathway has been found for $\rm Nb_3ICl_7$, we attempted a pure elemental gas-solid reaction method, as shown below:

\begin{equation}
3 \text{Nb} + \frac{7}{2} \text{Cl}_2 + \text{I} \rightarrow \text{Nb}_3\text{ICl}_7 + 13.932 \, \text{eV}
\end{equation}

\noindent This reaction exhibits a substantial formation enthalpy, suggesting a potentially high reaction rate that may be incompatible with the formation of an ordered kagome structure. Therefore, this pathway is provided for reference only. In summary, we have identified four potential pathways for synthesizing $\rm Nb_3FCl_7$, $\rm Nb_3BrCl_7$, and $\rm Nb_3ICl_7$, providing predictive guidance for experimental studies on breathing-type $\rm Nb_3XCl_7 $ systems.

\vspace{-0.2cm}
\section{IV. CONCLUSION}
\vspace{-0.3cm}

Based on the DFT+$U$ method and Boltzmann transport theory, we investigate the MIT and magnetic properties of kagome materials $\mathrm{Nb_3XCl_7}$ (X = F, Cl, Br, I), taking into account both the breathing effect and electron correlations. Through calculating the band structures with the HSE hybrid functional and the DFT+$U$ method, we determine that the band structures obtained with $U_{\text{eff}} = 2$ eV is in good agreement with experimental results. This conclusion is further corroborated by the calculated optical absorption spectra, JDOS, and magnetic ground state. Interestingly, the $\mathrm{Nb_3Cl_8}$ film exhibits an absorption rate of 51$\%$ near 3.04 eV in the violet light region, suggesting potential applications in novel optoelectronic devices.

We further elucidate the regulatory roles of electron correlation and the breathing effect in monolayer $\mathrm{Nb_3XCl_7}$ (X = F, Cl, Br, and I). Our findings reveal that monolayer $\mathrm{Nb_3Cl_8}$ undergoes a transition from a PM metal to a Mott insulator when $U_{\text{eff}} = 1$ eV. By substituting the Cl atom at the top of the Nb trimer with F, Br, and I, we subtly modulate the breathing strengths in $\mathrm{Nb_3XCl_7}$. To establish correspondence with the breathing strength characterized by DQMC method, we define the hopping ratio $t_{\text{out}}/t_{\text{in}}$ utilizing ELFs. The calculated $\sigma$ and magnetic properties indicate that the monolayer $\mathrm{Nb_3XCl_7}$ undergoes a transition from a PM metal to a Mott insulator around $t_{\text{out}}/t_{\text{in}} = 0.6674$. The combined influence of electron correlation and the breathing effect on $\mathrm{Nb_3XCl_7}$ remains consistent with theoretical predictions obtained from DQMC methods \cite{cpl_42_9_090712}. Through detailed analysis of the band structures and PDOS of $\mathrm{Nb_3XCl_7}$, we establish that stronger breathing strength corresponds to enhanced chemical precompression, which reduces the region of free electron gas between intercell Nb atoms and consequently suppresses electrical conductivity. Finally, we propose several viable synthesis pathways for $\mathrm{Nb_3FCl_7}$, $\mathrm{Nb_3BrCl_7}$, and $\mathrm{Nb_3ICl_7}$.

Our study establishes a practical framework for investigating the breathing effect in correlated kagome systems. Furthermore, it provides valuable insights into the microscopic mechanisms underlying MIT and magnetic properties in breathing kagome systems.

\vspace{-0.2cm}
\section{ACKNOWLEDGMENTS}
\vspace{-0.3cm}
This work was supported by NSFC (12474218) and Beijing Natural Science Foundation (No. 1242022 and 1252022). The numerical simulations in this work were performed at the HSCC of Beijing Normal University.

\bibliography{ref}

\begin{thebibliography}{59}%
\makeatletter
\providecommand \@ifxundefined [1]{%
 \@ifx{#1\undefined}
}%
\providecommand \@ifnum [1]{%
 \ifnum #1\expandafter \@firstoftwo
 \else \expandafter \@secondoftwo
 \fi
}%
\providecommand \@ifx [1]{%
 \ifx #1\expandafter \@firstoftwo
 \else \expandafter \@secondoftwo
 \fi
}%
\providecommand \natexlab [1]{#1}%
\providecommand \enquote  [1]{``#1''}%
\providecommand \bibnamefont  [1]{#1}%
\providecommand \bibfnamefont [1]{#1}%
\providecommand \citenamefont [1]{#1}%
\providecommand \href@noop [0]{\@secondoftwo}%
\providecommand \href [0]{\begingroup \@sanitize@url \@href}%
\providecommand \@href[1]{\@@startlink{#1}\@@href}%
\providecommand \@@href[1]{\endgroup#1\@@endlink}%
\providecommand \@sanitize@url [0]{\catcode `\\12\catcode `\$12\catcode
  `\&12\catcode `\#12\catcode `\^12\catcode `\_12\catcode `\%12\relax}%
\providecommand \@@startlink[1]{}%
\providecommand \@@endlink[0]{}%
\providecommand \url  [0]{\begingroup\@sanitize@url \@url }%
\providecommand \@url [1]{\endgroup\@href {#1}{\urlprefix }}%
\providecommand \urlprefix  [0]{URL }%
\providecommand \Eprint [0]{\href }%
\providecommand \doibase [0]{https://doi.org/}%
\providecommand \selectlanguage [0]{\@gobble}%
\providecommand \bibinfo  [0]{\@secondoftwo}%
\providecommand \bibfield  [0]{\@secondoftwo}%
\providecommand \translation [1]{[#1]}%
\providecommand \BibitemOpen [0]{}%
\providecommand \bibitemStop [0]{}%
\providecommand \bibitemNoStop [0]{.\EOS\space}%
\providecommand \EOS [0]{\spacefactor3000\relax}%
\providecommand \BibitemShut  [1]{\csname bibitem#1\endcsname}%
\let\auto@bib@innerbib\@empty
\bibitem [{\citenamefont {Cho}\ \emph {et~al.}(2021)\citenamefont {Cho},
  \citenamefont {Ma}, \citenamefont {Xia}, \citenamefont {Yang}, \citenamefont
  {Liu}, \citenamefont {Huang}, \citenamefont {Jiang}, \citenamefont {Lu},
  \citenamefont {Liu}, \citenamefont {Liu} \emph
  {et~al.}}]{PhysRevLett.127.236401}%
  \BibitemOpen
  \bibfield  {author} {\bibinfo {author} {\bibfnamefont {S.}~\bibnamefont
  {Cho}}, \bibinfo {author} {\bibfnamefont {H.}~\bibnamefont {Ma}}, \bibinfo
  {author} {\bibfnamefont {W.}~\bibnamefont {Xia}}, \bibinfo {author}
  {\bibfnamefont {Y.}~\bibnamefont {Yang}}, \bibinfo {author} {\bibfnamefont
  {Z.}~\bibnamefont {Liu}}, \bibinfo {author} {\bibfnamefont {Z.}~\bibnamefont
  {Huang}}, \bibinfo {author} {\bibfnamefont {Z.}~\bibnamefont {Jiang}},
  \bibinfo {author} {\bibfnamefont {X.}~\bibnamefont {Lu}}, \bibinfo {author}
  {\bibfnamefont {J.}~\bibnamefont {Liu}}, \bibinfo {author} {\bibfnamefont
  {Z.}~\bibnamefont {Liu}}, \emph {et~al.},\ }\href
  {https://link.aps.org/doi/10.1103/PhysRevLett.127.236401} {\bibfield
  {journal} {\bibinfo  {journal} {Phys. Rev. Lett.}\ }\textbf {\bibinfo
  {volume} {127}},\ \bibinfo {pages} {236401} (\bibinfo {year}
  {2021})}\BibitemShut {NoStop}%
\bibitem [{\citenamefont {Yi}\ \emph {et~al.}(2025)\citenamefont {Yi},
  \citenamefont {Liao}, \citenamefont {Wang}, \citenamefont {Ma}, \citenamefont
  {Liu}, \citenamefont {Teng}, \citenamefont {Gao}, \citenamefont {Dai},
  \citenamefont {Dai}, \citenamefont {Zhao} \emph
  {et~al.}}]{PhysRevLett.134.086902}%
  \BibitemOpen
  \bibfield  {author} {\bibinfo {author} {\bibfnamefont {S.}~\bibnamefont
  {Yi}}, \bibinfo {author} {\bibfnamefont {Z.}~\bibnamefont {Liao}}, \bibinfo
  {author} {\bibfnamefont {Q.}~\bibnamefont {Wang}}, \bibinfo {author}
  {\bibfnamefont {H.}~\bibnamefont {Ma}}, \bibinfo {author} {\bibfnamefont
  {J.}~\bibnamefont {Liu}}, \bibinfo {author} {\bibfnamefont {X.}~\bibnamefont
  {Teng}}, \bibinfo {author} {\bibfnamefont {B.}~\bibnamefont {Gao}}, \bibinfo
  {author} {\bibfnamefont {P.}~\bibnamefont {Dai}}, \bibinfo {author}
  {\bibfnamefont {Y.}~\bibnamefont {Dai}}, \bibinfo {author} {\bibfnamefont
  {J.}~\bibnamefont {Zhao}}, \emph {et~al.},\ }\href
  {https://link.aps.org/doi/10.1103/PhysRevLett.134.086902} {\bibfield
  {journal} {\bibinfo  {journal} {Phys. Rev. Lett.}\ }\textbf {\bibinfo
  {volume} {134}},\ \bibinfo {pages} {086902} (\bibinfo {year}
  {2025})}\BibitemShut {NoStop}%
\bibitem [{\citenamefont {Balents}(2010)}]{nature08917}%
  \BibitemOpen
  \bibfield  {author} {\bibinfo {author} {\bibfnamefont {L.}~\bibnamefont
  {Balents}},\ }\href {https://doi.org/10.1038/nature08917} {\bibfield
  {journal} {\bibinfo  {journal} {Nature}\ }\textbf {\bibinfo {volume} {464}},\
  \bibinfo {pages} {199} (\bibinfo {year} {2010})}\BibitemShut {NoStop}%
\bibitem [{\citenamefont {Han}\ \emph {et~al.}(2012)\citenamefont {Han},
  \citenamefont {Helton}, \citenamefont {Chu}, \citenamefont {Nocera},
  \citenamefont {Rodriguez-Rivera}, \citenamefont {Broholm},\ and\
  \citenamefont {Lee}}]{Nature.11659}%
  \BibitemOpen
  \bibfield  {author} {\bibinfo {author} {\bibfnamefont {T.-H.}\ \bibnamefont
  {Han}}, \bibinfo {author} {\bibfnamefont {J.~S.}\ \bibnamefont {Helton}},
  \bibinfo {author} {\bibfnamefont {S.}~\bibnamefont {Chu}}, \bibinfo {author}
  {\bibfnamefont {D.~G.}\ \bibnamefont {Nocera}}, \bibinfo {author}
  {\bibfnamefont {J.~A.}\ \bibnamefont {Rodriguez-Rivera}}, \bibinfo {author}
  {\bibfnamefont {C.}~\bibnamefont {Broholm}},\ and\ \bibinfo {author}
  {\bibfnamefont {Y.~S.}\ \bibnamefont {Lee}},\ }\href
  {https://doi.org/10.1038/nature11659} {\bibfield  {journal} {\bibinfo
  {journal} {Nature}\ }\textbf {\bibinfo {volume} {492}},\ \bibinfo {pages}
  {406} (\bibinfo {year} {2012})}\BibitemShut {NoStop}%
\bibitem [{\citenamefont {Fu}\ \emph {et~al.}(2015)\citenamefont {Fu},
  \citenamefont {Imai}, \citenamefont {Han},\ and\ \citenamefont
  {Lee}}]{science.aab2120}%
  \BibitemOpen
  \bibfield  {author} {\bibinfo {author} {\bibfnamefont {M.}~\bibnamefont
  {Fu}}, \bibinfo {author} {\bibfnamefont {T.}~\bibnamefont {Imai}}, \bibinfo
  {author} {\bibfnamefont {T.-H.}\ \bibnamefont {Han}},\ and\ \bibinfo {author}
  {\bibfnamefont {Y.~S.}\ \bibnamefont {Lee}},\ }\href
  {https://www.science.org/doi/abs/10.1126/science.aab2120} {\bibfield
  {journal} {\bibinfo  {journal} {Science}\ }\textbf {\bibinfo {volume}
  {350}},\ \bibinfo {pages} {655} (\bibinfo {year} {2015})}\BibitemShut
  {NoStop}%
\bibitem [{\citenamefont {Chen}\ \emph
  {et~al.}(2021{\natexlab{a}})\citenamefont {Chen}, \citenamefont {Wang},
  \citenamefont {Yin}, \citenamefont {Gu}, \citenamefont {Jiang}, \citenamefont
  {Tu}, \citenamefont {Gong}, \citenamefont {Uwatoko}, \citenamefont {Sun},
  \citenamefont {Lei}, \citenamefont {Hu}, \citenamefont {Cheng} \emph
  {et~al.}}]{PhysRevLett.126.247001}%
  \BibitemOpen
  \bibfield  {author} {\bibinfo {author} {\bibfnamefont {K.~Y.}\ \bibnamefont
  {Chen}}, \bibinfo {author} {\bibfnamefont {N.~N.}\ \bibnamefont {Wang}},
  \bibinfo {author} {\bibfnamefont {Q.~W.}\ \bibnamefont {Yin}}, \bibinfo
  {author} {\bibfnamefont {Y.~H.}\ \bibnamefont {Gu}}, \bibinfo {author}
  {\bibfnamefont {K.}~\bibnamefont {Jiang}}, \bibinfo {author} {\bibfnamefont
  {Z.~J.}\ \bibnamefont {Tu}}, \bibinfo {author} {\bibfnamefont {C.~S.}\
  \bibnamefont {Gong}}, \bibinfo {author} {\bibfnamefont {Y.}~\bibnamefont
  {Uwatoko}}, \bibinfo {author} {\bibfnamefont {J.~P.}\ \bibnamefont {Sun}},
  \bibinfo {author} {\bibfnamefont {H.~C.}\ \bibnamefont {Lei}}, \bibinfo
  {author} {\bibfnamefont {J.~P.}\ \bibnamefont {Hu}}, \bibinfo {author}
  {\bibfnamefont {J.-G.}\ \bibnamefont {Cheng}}, \emph {et~al.},\ }\href
  {https://link.aps.org/doi/10.1103/PhysRevLett.126.247001} {\bibfield
  {journal} {\bibinfo  {journal} {Phys. Rev. Lett.}\ }\textbf {\bibinfo
  {volume} {126}},\ \bibinfo {pages} {247001} (\bibinfo {year}
  {2021}{\natexlab{a}})}\BibitemShut {NoStop}%
\bibitem [{\citenamefont {Zhu}\ \emph {et~al.}(2023)\citenamefont {Zhu},
  \citenamefont {Yuan}, \citenamefont {Fang}, \citenamefont {Sun},\ and\
  \citenamefont {Wang}}]{ZHU2023157817}%
  \BibitemOpen
  \bibfield  {author} {\bibinfo {author} {\bibfnamefont {Y.}~\bibnamefont
  {Zhu}}, \bibinfo {author} {\bibfnamefont {J.-H.}\ \bibnamefont {Yuan}},
  \bibinfo {author} {\bibfnamefont {W.-Y.}\ \bibnamefont {Fang}}, \bibinfo
  {author} {\bibfnamefont {Z.-G.}\ \bibnamefont {Sun}},\ and\ \bibinfo {author}
  {\bibfnamefont {J.}~\bibnamefont {Wang}},\ }\href
  {https://www.sciencedirect.com/science/article/pii/S0169433223014964}
  {\bibfield  {journal} {\bibinfo  {journal} {Appl. Surf. Sci.}\ }\textbf
  {\bibinfo {volume} {636}},\ \bibinfo {pages} {157817} (\bibinfo {year}
  {2023})}\BibitemShut {NoStop}%
\bibitem [{\citenamefont {yu~Fang}\ \emph {et~al.}(2024)\citenamefont
  {yu~Fang}, \citenamefont {xiao Rao}, \citenamefont {Jin}, \citenamefont
  {an~Chen}, \citenamefont {fei Sheng},\ and\ \citenamefont
  {Bao}}]{FANG2024112725}%
  \BibitemOpen
  \bibfield  {author} {\bibinfo {author} {\bibfnamefont {W.}~\bibnamefont
  {yu~Fang}}, \bibinfo {author} {\bibfnamefont {X.}~\bibnamefont {xiao Rao}},
  \bibinfo {author} {\bibfnamefont {K.}~\bibnamefont {Jin}}, \bibinfo {author}
  {\bibfnamefont {S.}~\bibnamefont {an~Chen}}, \bibinfo {author} {\bibfnamefont
  {X.}~\bibnamefont {fei Sheng}},\ and\ \bibinfo {author} {\bibfnamefont
  {L.}~\bibnamefont {Bao}},\ }\href
  {https://www.sciencedirect.com/science/article/pii/S0042207X23009223}
  {\bibfield  {journal} {\bibinfo  {journal} {Vacuum}\ }\textbf {\bibinfo
  {volume} {219}},\ \bibinfo {pages} {112725} (\bibinfo {year}
  {2024})}\BibitemShut {NoStop}%
\bibitem [{\citenamefont {Yang}\ \emph {et~al.}(2024)\citenamefont {Yang},
  \citenamefont {Chen}, \citenamefont {Ma}, \citenamefont {Liang},\ and\
  \citenamefont {Ma}}]{Yang_2024}%
  \BibitemOpen
  \bibfield  {author} {\bibinfo {author} {\bibfnamefont {C.}~\bibnamefont
  {Yang}}, \bibinfo {author} {\bibfnamefont {C.}~\bibnamefont {Chen}}, \bibinfo
  {author} {\bibfnamefont {R.}~\bibnamefont {Ma}}, \bibinfo {author}
  {\bibfnamefont {Y.}~\bibnamefont {Liang}},\ and\ \bibinfo {author}
  {\bibfnamefont {T.}~\bibnamefont {Ma}},\ }\href
  {https://doi.org/10.1088/1674-1056/ad7578} {\bibfield  {journal} {\bibinfo
  {journal} {Chin. Phys. B}\ }\textbf {\bibinfo {volume} {33}},\ \bibinfo
  {pages} {107404} (\bibinfo {year} {2024})}\BibitemShut {NoStop}%
\bibitem [{\citenamefont {Yin}\ \emph {et~al.}(2020)\citenamefont {Yin},
  \citenamefont {Ma}, \citenamefont {Cochran}, \citenamefont {Xu},
  \citenamefont {Zhang}, \citenamefont {Tien}, \citenamefont {Shumiya},
  \citenamefont {Cheng}, \citenamefont {Jiang}, \citenamefont {Lian} \emph
  {et~al.}}]{Nature.s41586-020-2482-7}%
  \BibitemOpen
  \bibfield  {author} {\bibinfo {author} {\bibfnamefont {J.-X.}\ \bibnamefont
  {Yin}}, \bibinfo {author} {\bibfnamefont {W.}~\bibnamefont {Ma}}, \bibinfo
  {author} {\bibfnamefont {T.~A.}\ \bibnamefont {Cochran}}, \bibinfo {author}
  {\bibfnamefont {X.}~\bibnamefont {Xu}}, \bibinfo {author} {\bibfnamefont
  {S.~S.}\ \bibnamefont {Zhang}}, \bibinfo {author} {\bibfnamefont {H.-J.}\
  \bibnamefont {Tien}}, \bibinfo {author} {\bibfnamefont {N.}~\bibnamefont
  {Shumiya}}, \bibinfo {author} {\bibfnamefont {G.}~\bibnamefont {Cheng}},
  \bibinfo {author} {\bibfnamefont {K.}~\bibnamefont {Jiang}}, \bibinfo
  {author} {\bibfnamefont {B.}~\bibnamefont {Lian}}, \emph {et~al.},\ }\href
  {https://doi.org/10.1038/s41586-020-2482-7} {\bibfield  {journal} {\bibinfo
  {journal} {Nature}\ }\textbf {\bibinfo {volume} {583}},\ \bibinfo {pages}
  {533} (\bibinfo {year} {2020})}\BibitemShut {NoStop}%
\bibitem [{\citenamefont {Tang}\ \emph {et~al.}(2011)\citenamefont {Tang},
  \citenamefont {Mei},\ and\ \citenamefont {Wen}}]{PhysRevLett.106.236802}%
  \BibitemOpen
  \bibfield  {author} {\bibinfo {author} {\bibfnamefont {E.}~\bibnamefont
  {Tang}}, \bibinfo {author} {\bibfnamefont {J.-W.}\ \bibnamefont {Mei}},\ and\
  \bibinfo {author} {\bibfnamefont {X.-G.}\ \bibnamefont {Wen}},\ }\href
  {https://link.aps.org/doi/10.1103/PhysRevLett.106.236802} {\bibfield
  {journal} {\bibinfo  {journal} {Phys. Rev. Lett.}\ }\textbf {\bibinfo
  {volume} {106}},\ \bibinfo {pages} {236802} (\bibinfo {year}
  {2011})}\BibitemShut {NoStop}%
\bibitem [{\citenamefont {Ye}\ \emph {et~al.}(2018)\citenamefont {Ye},
  \citenamefont {Kang}, \citenamefont {Liu}, \citenamefont {von Cube},
  \citenamefont {Wicker}, \citenamefont {Suzuki}, \citenamefont {Jozwiak},
  \citenamefont {Bostwick}, \citenamefont {Rotenberg}, \citenamefont {Bell}
  \emph {et~al.}}]{nature25987}%
  \BibitemOpen
  \bibfield  {author} {\bibinfo {author} {\bibfnamefont {L.}~\bibnamefont
  {Ye}}, \bibinfo {author} {\bibfnamefont {M.}~\bibnamefont {Kang}}, \bibinfo
  {author} {\bibfnamefont {J.}~\bibnamefont {Liu}}, \bibinfo {author}
  {\bibfnamefont {F.}~\bibnamefont {von Cube}}, \bibinfo {author}
  {\bibfnamefont {C.~R.}\ \bibnamefont {Wicker}}, \bibinfo {author}
  {\bibfnamefont {T.}~\bibnamefont {Suzuki}}, \bibinfo {author} {\bibfnamefont
  {C.}~\bibnamefont {Jozwiak}}, \bibinfo {author} {\bibfnamefont
  {A.}~\bibnamefont {Bostwick}}, \bibinfo {author} {\bibfnamefont
  {E.}~\bibnamefont {Rotenberg}}, \bibinfo {author} {\bibfnamefont {D.~C.}\
  \bibnamefont {Bell}}, \emph {et~al.},\ }\href
  {https://doi.org/10.1038/nature25987} {\bibfield  {journal} {\bibinfo
  {journal} {Nature}\ }\textbf {\bibinfo {volume} {555}},\ \bibinfo {pages}
  {638} (\bibinfo {year} {2018})}\BibitemShut {NoStop}%
\bibitem [{\citenamefont {Liu}\ \emph {et~al.}(2021)\citenamefont {Liu},
  \citenamefont {Zhao}, \citenamefont {Yin}, \citenamefont {Gong},
  \citenamefont {Tu}, \citenamefont {Li}, \citenamefont {Song}, \citenamefont
  {Liu}, \citenamefont {Shen}, \citenamefont {Huang} \emph
  {et~al.}}]{PhysRevX.11.041010}%
  \BibitemOpen
  \bibfield  {author} {\bibinfo {author} {\bibfnamefont {Z.}~\bibnamefont
  {Liu}}, \bibinfo {author} {\bibfnamefont {N.}~\bibnamefont {Zhao}}, \bibinfo
  {author} {\bibfnamefont {Q.}~\bibnamefont {Yin}}, \bibinfo {author}
  {\bibfnamefont {C.}~\bibnamefont {Gong}}, \bibinfo {author} {\bibfnamefont
  {Z.}~\bibnamefont {Tu}}, \bibinfo {author} {\bibfnamefont {M.}~\bibnamefont
  {Li}}, \bibinfo {author} {\bibfnamefont {W.}~\bibnamefont {Song}}, \bibinfo
  {author} {\bibfnamefont {Z.}~\bibnamefont {Liu}}, \bibinfo {author}
  {\bibfnamefont {D.}~\bibnamefont {Shen}}, \bibinfo {author} {\bibfnamefont
  {Y.}~\bibnamefont {Huang}}, \emph {et~al.},\ }\href
  {https://link.aps.org/doi/10.1103/PhysRevX.11.041010} {\bibfield  {journal}
  {\bibinfo  {journal} {Phys. Rev. X}\ }\textbf {\bibinfo {volume} {11}},\
  \bibinfo {pages} {041010} (\bibinfo {year} {2021})}\BibitemShut {NoStop}%
\bibitem [{\citenamefont {Chen}\ \emph
  {et~al.}(2021{\natexlab{b}})\citenamefont {Chen}, \citenamefont {Yang},
  \citenamefont {Hu}, \citenamefont {Zhao}, \citenamefont {Yuan}, \citenamefont
  {Xing}, \citenamefont {Qian}, \citenamefont {Huang}, \citenamefont {Li},
  \citenamefont {Ye} \emph {et~al.}}]{Nature.s41586-021-03983-5}%
  \BibitemOpen
  \bibfield  {author} {\bibinfo {author} {\bibfnamefont {H.}~\bibnamefont
  {Chen}}, \bibinfo {author} {\bibfnamefont {H.}~\bibnamefont {Yang}}, \bibinfo
  {author} {\bibfnamefont {B.}~\bibnamefont {Hu}}, \bibinfo {author}
  {\bibfnamefont {Z.}~\bibnamefont {Zhao}}, \bibinfo {author} {\bibfnamefont
  {J.}~\bibnamefont {Yuan}}, \bibinfo {author} {\bibfnamefont {Y.}~\bibnamefont
  {Xing}}, \bibinfo {author} {\bibfnamefont {G.}~\bibnamefont {Qian}}, \bibinfo
  {author} {\bibfnamefont {Z.}~\bibnamefont {Huang}}, \bibinfo {author}
  {\bibfnamefont {G.}~\bibnamefont {Li}}, \bibinfo {author} {\bibfnamefont
  {Y.}~\bibnamefont {Ye}}, \emph {et~al.},\ }\href
  {https://doi.org/10.1038/s41586-021-03983-5} {\bibfield  {journal} {\bibinfo
  {journal} {Nature}\ }\textbf {\bibinfo {volume} {599}},\ \bibinfo {pages}
  {222} (\bibinfo {year} {2021}{\natexlab{b}})}\BibitemShut {NoStop}%
\bibitem [{\citenamefont {Jiang}\ \emph {et~al.}(2021)\citenamefont {Jiang},
  \citenamefont {Yin}, \citenamefont {Denner}, \citenamefont {Shumiya},
  \citenamefont {Ortiz}, \citenamefont {Xu}, \citenamefont {Guguchia},
  \citenamefont {He}, \citenamefont {Hossain}, \citenamefont {Liu} \emph
  {et~al.}}]{Nat.Mater.s41563-021-01034-y}%
  \BibitemOpen
  \bibfield  {author} {\bibinfo {author} {\bibfnamefont {Y.-X.}\ \bibnamefont
  {Jiang}}, \bibinfo {author} {\bibfnamefont {J.-X.}\ \bibnamefont {Yin}},
  \bibinfo {author} {\bibfnamefont {M.~M.}\ \bibnamefont {Denner}}, \bibinfo
  {author} {\bibfnamefont {N.}~\bibnamefont {Shumiya}}, \bibinfo {author}
  {\bibfnamefont {B.~R.}\ \bibnamefont {Ortiz}}, \bibinfo {author}
  {\bibfnamefont {G.}~\bibnamefont {Xu}}, \bibinfo {author} {\bibfnamefont
  {Z.}~\bibnamefont {Guguchia}}, \bibinfo {author} {\bibfnamefont
  {J.}~\bibnamefont {He}}, \bibinfo {author} {\bibfnamefont {M.~S.}\
  \bibnamefont {Hossain}}, \bibinfo {author} {\bibfnamefont {X.}~\bibnamefont
  {Liu}}, \emph {et~al.},\ }\href {https://doi.org/10.1038/s41563-021-01034-y}
  {\bibfield  {journal} {\bibinfo  {journal} {Nat. Mater.}\ }\textbf {\bibinfo
  {volume} {20}},\ \bibinfo {pages} {1353} (\bibinfo {year}
  {2021})}\BibitemShut {NoStop}%
\bibitem [{\citenamefont {Jiang}\ \emph {et~al.}(2022)\citenamefont {Jiang},
  \citenamefont {Wu}, \citenamefont {Yin}, \citenamefont {Wang}, \citenamefont
  {Hasan}, \citenamefont {Wilson}, \citenamefont {Chen},\ and\ \citenamefont
  {Hu}}]{10.1093/nsr/nwac199}%
  \BibitemOpen
  \bibfield  {author} {\bibinfo {author} {\bibfnamefont {K.}~\bibnamefont
  {Jiang}}, \bibinfo {author} {\bibfnamefont {T.}~\bibnamefont {Wu}}, \bibinfo
  {author} {\bibfnamefont {J.-X.}\ \bibnamefont {Yin}}, \bibinfo {author}
  {\bibfnamefont {Z.}~\bibnamefont {Wang}}, \bibinfo {author} {\bibfnamefont
  {M.~Z.}\ \bibnamefont {Hasan}}, \bibinfo {author} {\bibfnamefont {S.~D.}\
  \bibnamefont {Wilson}}, \bibinfo {author} {\bibfnamefont {X.}~\bibnamefont
  {Chen}},\ and\ \bibinfo {author} {\bibfnamefont {J.}~\bibnamefont {Hu}},\
  }\href {https://doi.org/10.1093/nsr/nwac199} {\bibfield  {journal} {\bibinfo
  {journal} {Natl. Sci. Rev.}\ }\textbf {\bibinfo {volume} {10}},\ \bibinfo
  {pages} {nwac199} (\bibinfo {year} {2022})}\BibitemShut {NoStop}%
\bibitem [{\citenamefont {Lyalin}\ \emph {et~al.}(2021)\citenamefont {Lyalin},
  \citenamefont {Cheng},\ and\ \citenamefont
  {Kawakami}}]{acs.nanolett.1c02270}%
  \BibitemOpen
  \bibfield  {author} {\bibinfo {author} {\bibfnamefont {I.}~\bibnamefont
  {Lyalin}}, \bibinfo {author} {\bibfnamefont {S.}~\bibnamefont {Cheng}},\ and\
  \bibinfo {author} {\bibfnamefont {R.~K.}\ \bibnamefont {Kawakami}},\ }\href
  {https://doi.org/10.1021/acs.nanolett.1c02270} {\bibfield  {journal}
  {\bibinfo  {journal} {Nano Lett.}\ }\textbf {\bibinfo {volume} {21}},\
  \bibinfo {pages} {6975} (\bibinfo {year} {2021})}\BibitemShut {NoStop}%
\bibitem [{\citenamefont {Guin}\ \emph {et~al.}(2019)\citenamefont {Guin},
  \citenamefont {Vir}, \citenamefont {Zhang}, \citenamefont {Kumar},
  \citenamefont {Watzman}, \citenamefont {Fu}, \citenamefont {Liu},
  \citenamefont {Manna}, \citenamefont {Schnelle}, \citenamefont {Gooth} \emph
  {et~al.}}]{adma.201806622}%
  \BibitemOpen
  \bibfield  {author} {\bibinfo {author} {\bibfnamefont {S.~N.}\ \bibnamefont
  {Guin}}, \bibinfo {author} {\bibfnamefont {P.}~\bibnamefont {Vir}}, \bibinfo
  {author} {\bibfnamefont {Y.}~\bibnamefont {Zhang}}, \bibinfo {author}
  {\bibfnamefont {N.}~\bibnamefont {Kumar}}, \bibinfo {author} {\bibfnamefont
  {S.~J.}\ \bibnamefont {Watzman}}, \bibinfo {author} {\bibfnamefont
  {C.}~\bibnamefont {Fu}}, \bibinfo {author} {\bibfnamefont {E.}~\bibnamefont
  {Liu}}, \bibinfo {author} {\bibfnamefont {K.}~\bibnamefont {Manna}}, \bibinfo
  {author} {\bibfnamefont {W.}~\bibnamefont {Schnelle}}, \bibinfo {author}
  {\bibfnamefont {J.}~\bibnamefont {Gooth}}, \emph {et~al.},\ }\href
  {https://advanced.onlinelibrary.wiley.com/doi/abs/10.1002/adma.201806622}
  {\bibfield  {journal} {\bibinfo  {journal} {Adv. Mater.}\ }\textbf {\bibinfo
  {volume} {31}},\ \bibinfo {pages} {1806622} (\bibinfo {year}
  {2019})}\BibitemShut {NoStop}%
\bibitem [{\citenamefont {Zhang}\ \emph {et~al.}(2022)\citenamefont {Zhang},
  \citenamefont {Yilmaz}, \citenamefont {Vescovo}, \citenamefont {Li},
  \citenamefont {Moore}, \citenamefont {Lee}, \citenamefont {Miao},
  \citenamefont {Murakami},\ and\ \citenamefont
  {McGuire}}]{s41524-022-00838-z}%
  \BibitemOpen
  \bibfield  {author} {\bibinfo {author} {\bibfnamefont {T.}~\bibnamefont
  {Zhang}}, \bibinfo {author} {\bibfnamefont {T.}~\bibnamefont {Yilmaz}},
  \bibinfo {author} {\bibfnamefont {E.}~\bibnamefont {Vescovo}}, \bibinfo
  {author} {\bibfnamefont {H.~X.}\ \bibnamefont {Li}}, \bibinfo {author}
  {\bibfnamefont {R.~G.}\ \bibnamefont {Moore}}, \bibinfo {author}
  {\bibfnamefont {H.~N.}\ \bibnamefont {Lee}}, \bibinfo {author} {\bibfnamefont
  {H.}~\bibnamefont {Miao}}, \bibinfo {author} {\bibfnamefont {S.}~\bibnamefont
  {Murakami}},\ and\ \bibinfo {author} {\bibfnamefont {M.~A.}\ \bibnamefont
  {McGuire}},\ }\href {https://doi.org/10.1038/s41524-022-00838-z} {\bibfield
  {journal} {\bibinfo  {journal} {npj Comput. Mater.}\ }\textbf {\bibinfo
  {volume} {8}},\ \bibinfo {pages} {155} (\bibinfo {year} {2022})}\BibitemShut
  {NoStop}%
\bibitem [{\citenamefont {Kang}\ \emph {et~al.}(2020)\citenamefont {Kang},
  \citenamefont {Fang}, \citenamefont {Ye}, \citenamefont {Po}, \citenamefont
  {Denlinger}, \citenamefont {Jozwiak}, \citenamefont {Bostwick}, \citenamefont
  {Rotenberg}, \citenamefont {Kaxiras}, \citenamefont {Checkelsky} \emph
  {et~al.}}]{s41467-020-17465-1}%
  \BibitemOpen
  \bibfield  {author} {\bibinfo {author} {\bibfnamefont {M.}~\bibnamefont
  {Kang}}, \bibinfo {author} {\bibfnamefont {S.}~\bibnamefont {Fang}}, \bibinfo
  {author} {\bibfnamefont {L.}~\bibnamefont {Ye}}, \bibinfo {author}
  {\bibfnamefont {H.~C.}\ \bibnamefont {Po}}, \bibinfo {author} {\bibfnamefont
  {J.}~\bibnamefont {Denlinger}}, \bibinfo {author} {\bibfnamefont
  {C.}~\bibnamefont {Jozwiak}}, \bibinfo {author} {\bibfnamefont
  {A.}~\bibnamefont {Bostwick}}, \bibinfo {author} {\bibfnamefont
  {E.}~\bibnamefont {Rotenberg}}, \bibinfo {author} {\bibfnamefont
  {E.}~\bibnamefont {Kaxiras}}, \bibinfo {author} {\bibfnamefont {J.~G.}\
  \bibnamefont {Checkelsky}}, \emph {et~al.},\ }\href
  {https://doi.org/10.1038/s41467-020-17465-1} {\bibfield  {journal} {\bibinfo
  {journal} {Nat. Commun.}\ }\textbf {\bibinfo {volume} {11}},\ \bibinfo
  {pages} {4004} (\bibinfo {year} {2020})}\BibitemShut {NoStop}%
\bibitem [{\citenamefont {Yin}\ \emph {et~al.}(2022)\citenamefont {Yin},
  \citenamefont {Jiang}, \citenamefont {Teng}, \citenamefont {Hossain},
  \citenamefont {Mardanya}, \citenamefont {Chang}, \citenamefont {Ye},
  \citenamefont {Xu}, \citenamefont {Denner}, \citenamefont {Neupert} \emph
  {et~al.}}]{PhysRevLett.129.166401}%
  \BibitemOpen
  \bibfield  {author} {\bibinfo {author} {\bibfnamefont {J.-X.}\ \bibnamefont
  {Yin}}, \bibinfo {author} {\bibfnamefont {Y.-X.}\ \bibnamefont {Jiang}},
  \bibinfo {author} {\bibfnamefont {X.}~\bibnamefont {Teng}}, \bibinfo {author}
  {\bibfnamefont {M.~S.}\ \bibnamefont {Hossain}}, \bibinfo {author}
  {\bibfnamefont {S.}~\bibnamefont {Mardanya}}, \bibinfo {author}
  {\bibfnamefont {T.-R.}\ \bibnamefont {Chang}}, \bibinfo {author}
  {\bibfnamefont {Z.}~\bibnamefont {Ye}}, \bibinfo {author} {\bibfnamefont
  {G.}~\bibnamefont {Xu}}, \bibinfo {author} {\bibfnamefont {M.~M.}\
  \bibnamefont {Denner}}, \bibinfo {author} {\bibfnamefont {T.}~\bibnamefont
  {Neupert}}, \emph {et~al.},\ }\href
  {https://link.aps.org/doi/10.1103/PhysRevLett.129.166401} {\bibfield
  {journal} {\bibinfo  {journal} {Phys. Rev. Lett.}\ }\textbf {\bibinfo
  {volume} {129}},\ \bibinfo {pages} {166401} (\bibinfo {year}
  {2022})}\BibitemShut {NoStop}%
\bibitem [{\citenamefont {Xu}\ \emph {et~al.}(2022)\citenamefont {Xu},
  \citenamefont {Yin}, \citenamefont {Ma}, \citenamefont {Tien}, \citenamefont
  {Qiang}, \citenamefont {Reddy}, \citenamefont {Zhou}, \citenamefont {Shen},
  \citenamefont {Lu}, \citenamefont {Chang} \emph {et~al.}}]{Xu2022}%
  \BibitemOpen
  \bibfield  {author} {\bibinfo {author} {\bibfnamefont {X.}~\bibnamefont
  {Xu}}, \bibinfo {author} {\bibfnamefont {J.-X.}\ \bibnamefont {Yin}},
  \bibinfo {author} {\bibfnamefont {W.}~\bibnamefont {Ma}}, \bibinfo {author}
  {\bibfnamefont {H.-J.}\ \bibnamefont {Tien}}, \bibinfo {author}
  {\bibfnamefont {X.-B.}\ \bibnamefont {Qiang}}, \bibinfo {author}
  {\bibfnamefont {P.~V.~S.}\ \bibnamefont {Reddy}}, \bibinfo {author}
  {\bibfnamefont {H.}~\bibnamefont {Zhou}}, \bibinfo {author} {\bibfnamefont
  {J.}~\bibnamefont {Shen}}, \bibinfo {author} {\bibfnamefont {H.-Z.}\
  \bibnamefont {Lu}}, \bibinfo {author} {\bibfnamefont {T.-R.}\ \bibnamefont
  {Chang}}, \emph {et~al.},\ }\href
  {https://doi.org/10.1038/s41467-022-28796-6} {\bibfield  {journal} {\bibinfo
  {journal} {Nat. Commun.}\ }\textbf {\bibinfo {volume} {13}},\ \bibinfo
  {pages} {1197} (\bibinfo {year} {2022})}\BibitemShut {NoStop}%
\bibitem [{\citenamefont {Sun}\ \emph {et~al.}(2022)\citenamefont {Sun},
  \citenamefont {Zhou}, \citenamefont {Wang}, \citenamefont {Kumar},
  \citenamefont {Geng}, \citenamefont {Yue}, \citenamefont {Han}, \citenamefont
  {Haraguchi}, \citenamefont {Shimada}, \citenamefont {Cheng} \emph
  {et~al.}}]{acs.nanolett.2c00778}%
  \BibitemOpen
  \bibfield  {author} {\bibinfo {author} {\bibfnamefont {Z.}~\bibnamefont
  {Sun}}, \bibinfo {author} {\bibfnamefont {H.}~\bibnamefont {Zhou}}, \bibinfo
  {author} {\bibfnamefont {C.}~\bibnamefont {Wang}}, \bibinfo {author}
  {\bibfnamefont {S.}~\bibnamefont {Kumar}}, \bibinfo {author} {\bibfnamefont
  {D.}~\bibnamefont {Geng}}, \bibinfo {author} {\bibfnamefont {S.}~\bibnamefont
  {Yue}}, \bibinfo {author} {\bibfnamefont {X.}~\bibnamefont {Han}}, \bibinfo
  {author} {\bibfnamefont {Y.}~\bibnamefont {Haraguchi}}, \bibinfo {author}
  {\bibfnamefont {K.}~\bibnamefont {Shimada}}, \bibinfo {author} {\bibfnamefont
  {P.}~\bibnamefont {Cheng}}, \emph {et~al.},\ }\href
  {https://doi.org/10.1021/acs.nanolett.2c00778} {\bibfield  {journal}
  {\bibinfo  {journal} {Nano Lett.}\ }\textbf {\bibinfo {volume} {22}},\
  \bibinfo {pages} {4596} (\bibinfo {year} {2022})}\BibitemShut {NoStop}%
\bibitem [{\citenamefont {Zhang}\ \emph {et~al.}(2023)\citenamefont {Zhang},
  \citenamefont {Shi}, \citenamefont {Jiang}, \citenamefont {Yang},
  \citenamefont {Zhang}, \citenamefont {Meng}, \citenamefont {Hu},
  \citenamefont {Liu}, \citenamefont {Cheng}, \citenamefont {Xie} \emph
  {et~al.}}]{adma.202301790}%
  \BibitemOpen
  \bibfield  {author} {\bibinfo {author} {\bibfnamefont {H.}~\bibnamefont
  {Zhang}}, \bibinfo {author} {\bibfnamefont {Z.}~\bibnamefont {Shi}}, \bibinfo
  {author} {\bibfnamefont {Z.}~\bibnamefont {Jiang}}, \bibinfo {author}
  {\bibfnamefont {M.}~\bibnamefont {Yang}}, \bibinfo {author} {\bibfnamefont
  {J.}~\bibnamefont {Zhang}}, \bibinfo {author} {\bibfnamefont
  {Z.}~\bibnamefont {Meng}}, \bibinfo {author} {\bibfnamefont {T.}~\bibnamefont
  {Hu}}, \bibinfo {author} {\bibfnamefont {F.}~\bibnamefont {Liu}}, \bibinfo
  {author} {\bibfnamefont {L.}~\bibnamefont {Cheng}}, \bibinfo {author}
  {\bibfnamefont {Y.}~\bibnamefont {Xie}}, \emph {et~al.},\ }\href
  {https://doi.org/10.1002/adma.202301790} {\bibfield  {journal} {\bibinfo
  {journal} {Adv. Mater.}\ }\textbf {\bibinfo {volume} {35}},\ \bibinfo {pages}
  {2301790} (\bibinfo {year} {2023})}\BibitemShut {NoStop}%
\bibitem [{\citenamefont {Hong}\ \emph {et~al.}(2025)\citenamefont {Hong},
  \citenamefont {Dai}, \citenamefont {Hu}, \citenamefont {Li},\ and\
  \citenamefont {Wu}}]{k3fv-9hss}%
  \BibitemOpen
  \bibfield  {author} {\bibinfo {author} {\bibfnamefont {M.}~\bibnamefont
  {Hong}}, \bibinfo {author} {\bibfnamefont {L.}~\bibnamefont {Dai}}, \bibinfo
  {author} {\bibfnamefont {H.}~\bibnamefont {Hu}}, \bibinfo {author}
  {\bibfnamefont {C.}~\bibnamefont {Li}},\ and\ \bibinfo {author}
  {\bibfnamefont {M.}~\bibnamefont {Wu}},\ }\href
  {https://link.aps.org/doi/10.1103/k3fv-9hss} {\bibfield  {journal} {\bibinfo
  {journal} {Phys. Rev. B}\ }\textbf {\bibinfo {volume} {111}},\ \bibinfo
  {pages} {224107} (\bibinfo {year} {2025})}\BibitemShut {NoStop}%
\bibitem [{\citenamefont {Teng}\ \emph {et~al.}(2022)\citenamefont {Teng},
  \citenamefont {Chen}, \citenamefont {Ye}, \citenamefont {Rosenberg},
  \citenamefont {Liu}, \citenamefont {Yin}, \citenamefont {Jiang},
  \citenamefont {Oh}, \citenamefont {Hasan}, \citenamefont {Neubauer} \emph
  {et~al.}}]{Nature.022.05034}%
  \BibitemOpen
  \bibfield  {author} {\bibinfo {author} {\bibfnamefont {X.}~\bibnamefont
  {Teng}}, \bibinfo {author} {\bibfnamefont {L.}~\bibnamefont {Chen}}, \bibinfo
  {author} {\bibfnamefont {F.}~\bibnamefont {Ye}}, \bibinfo {author}
  {\bibfnamefont {E.}~\bibnamefont {Rosenberg}}, \bibinfo {author}
  {\bibfnamefont {Z.}~\bibnamefont {Liu}}, \bibinfo {author} {\bibfnamefont
  {J.-X.}\ \bibnamefont {Yin}}, \bibinfo {author} {\bibfnamefont {Y.-X.}\
  \bibnamefont {Jiang}}, \bibinfo {author} {\bibfnamefont {J.~S.}\ \bibnamefont
  {Oh}}, \bibinfo {author} {\bibfnamefont {M.~Z.}\ \bibnamefont {Hasan}},
  \bibinfo {author} {\bibfnamefont {K.~J.}\ \bibnamefont {Neubauer}}, \emph
  {et~al.},\ }\href {https://doi.org/10.1038/s41586-022-05034-z} {\bibfield
  {journal} {\bibinfo  {journal} {Nature}\ }\textbf {\bibinfo {volume} {609}},\
  \bibinfo {pages} {490} (\bibinfo {year} {2022})}\BibitemShut {NoStop}%
\bibitem [{\citenamefont {Wang}\ \emph
  {et~al.}(2021{\natexlab{a}})\citenamefont {Wang}, \citenamefont {Neubauer},
  \citenamefont {Duan}, \citenamefont {Yin}, \citenamefont {Fujitsu},
  \citenamefont {Hosono}, \citenamefont {Ye}, \citenamefont {Zhang},
  \citenamefont {Chi}, \citenamefont {Krycka} \emph
  {et~al.}}]{PhysRevB.103.014416}%
  \BibitemOpen
  \bibfield  {author} {\bibinfo {author} {\bibfnamefont {Q.}~\bibnamefont
  {Wang}}, \bibinfo {author} {\bibfnamefont {K.~J.}\ \bibnamefont {Neubauer}},
  \bibinfo {author} {\bibfnamefont {C.}~\bibnamefont {Duan}}, \bibinfo {author}
  {\bibfnamefont {Q.}~\bibnamefont {Yin}}, \bibinfo {author} {\bibfnamefont
  {S.}~\bibnamefont {Fujitsu}}, \bibinfo {author} {\bibfnamefont
  {H.}~\bibnamefont {Hosono}}, \bibinfo {author} {\bibfnamefont
  {F.}~\bibnamefont {Ye}}, \bibinfo {author} {\bibfnamefont {R.}~\bibnamefont
  {Zhang}}, \bibinfo {author} {\bibfnamefont {S.}~\bibnamefont {Chi}}, \bibinfo
  {author} {\bibfnamefont {K.}~\bibnamefont {Krycka}}, \emph {et~al.},\ }\href
  {https://link.aps.org/doi/10.1103/PhysRevB.103.014416} {\bibfield  {journal}
  {\bibinfo  {journal} {Phys. Rev. B}\ }\textbf {\bibinfo {volume} {103}},\
  \bibinfo {pages} {014416} (\bibinfo {year} {2021}{\natexlab{a}})}\BibitemShut
  {NoStop}%
\bibitem [{\citenamefont {Wu}\ \emph {et~al.}(2025)\citenamefont {Wu},
  \citenamefont {Xu}, \citenamefont {Wang}, \citenamefont {Lin}, \citenamefont
  {Cao},\ and\ \citenamefont {Cao}}]{Nat.Commun.16.1375}%
  \BibitemOpen
  \bibfield  {author} {\bibinfo {author} {\bibfnamefont {S.}~\bibnamefont
  {Wu}}, \bibinfo {author} {\bibfnamefont {C.}~\bibnamefont {Xu}}, \bibinfo
  {author} {\bibfnamefont {X.}~\bibnamefont {Wang}}, \bibinfo {author}
  {\bibfnamefont {H.-Q.}\ \bibnamefont {Lin}}, \bibinfo {author} {\bibfnamefont
  {C.}~\bibnamefont {Cao}},\ and\ \bibinfo {author} {\bibfnamefont {G.-H.}\
  \bibnamefont {Cao}},\ }\href {https://doi.org/10.1038/s41467-025-56582-7}
  {\bibfield  {journal} {\bibinfo  {journal} {Nat. Commun.}\ }\textbf {\bibinfo
  {volume} {16}},\ \bibinfo {pages} {1375} (\bibinfo {year}
  {2025})}\BibitemShut {NoStop}%
\bibitem [{\citenamefont {Chen}\ \emph {et~al.}(2024)\citenamefont {Chen},
  \citenamefont {Zhou}, \citenamefont {Zhang}, \citenamefont {Ji},
  \citenamefont {Liao}, \citenamefont {Ji}, \citenamefont {Li}, \citenamefont
  {Guo}, \citenamefont {Shen}, \citenamefont {Yu} \emph
  {et~al.}}]{CommunMater.024.00513}%
  \BibitemOpen
  \bibfield  {author} {\bibinfo {author} {\bibfnamefont {L.}~\bibnamefont
  {Chen}}, \bibinfo {author} {\bibfnamefont {Y.}~\bibnamefont {Zhou}}, \bibinfo
  {author} {\bibfnamefont {H.}~\bibnamefont {Zhang}}, \bibinfo {author}
  {\bibfnamefont {X.}~\bibnamefont {Ji}}, \bibinfo {author} {\bibfnamefont
  {K.}~\bibnamefont {Liao}}, \bibinfo {author} {\bibfnamefont {Y.}~\bibnamefont
  {Ji}}, \bibinfo {author} {\bibfnamefont {Y.}~\bibnamefont {Li}}, \bibinfo
  {author} {\bibfnamefont {Z.}~\bibnamefont {Guo}}, \bibinfo {author}
  {\bibfnamefont {X.}~\bibnamefont {Shen}}, \bibinfo {author} {\bibfnamefont
  {R.}~\bibnamefont {Yu}}, \emph {et~al.},\ }\href
  {https://doi.org/10.1038/s43246-024-00513-4} {\bibfield  {journal} {\bibinfo
  {journal} {Commun. Mater.}\ }\textbf {\bibinfo {volume} {5}},\ \bibinfo
  {pages} {73} (\bibinfo {year} {2024})}\BibitemShut {NoStop}%
\bibitem [{\citenamefont {Li}\ \emph {et~al.}(2021)\citenamefont {Li},
  \citenamefont {Liu}, \citenamefont {Zhao}, \citenamefont {Hu},\ and\
  \citenamefont {Ren}}]{PhysRevB.104.L060405}%
  \BibitemOpen
  \bibfield  {author} {\bibinfo {author} {\bibfnamefont {Y.}~\bibnamefont
  {Li}}, \bibinfo {author} {\bibfnamefont {C.}~\bibnamefont {Liu}}, \bibinfo
  {author} {\bibfnamefont {G.-D.}\ \bibnamefont {Zhao}}, \bibinfo {author}
  {\bibfnamefont {T.}~\bibnamefont {Hu}},\ and\ \bibinfo {author}
  {\bibfnamefont {W.}~\bibnamefont {Ren}},\ }\href
  {https://link.aps.org/doi/10.1103/PhysRevB.104.L060405} {\bibfield  {journal}
  {\bibinfo  {journal} {Phys. Rev. B}\ }\textbf {\bibinfo {volume} {104}},\
  \bibinfo {pages} {L060405} (\bibinfo {year} {2021})}\BibitemShut {NoStop}%
\bibitem [{\citenamefont {Nakamura}\ \emph {et~al.}()\citenamefont {Nakamura},
  \citenamefont {Chudo},\ and\ \citenamefont {Shiga}}]{Nakamura_2005}%
  \BibitemOpen
  \bibfield  {author} {\bibinfo {author} {\bibfnamefont {H.}~\bibnamefont
  {Nakamura}}, \bibinfo {author} {\bibfnamefont {H.}~\bibnamefont {Chudo}},\
  and\ \bibinfo {author} {\bibfnamefont {M.}~\bibnamefont {Shiga}},\ }\href
  {https://doi.org/10.1088/0953-8984/17/38/007} {\bibfield  {journal} {\bibinfo
   {journal} {J. Phys.: Condens. Matter}\ }\textbf {\bibinfo {volume} {17}},\
  \bibinfo {pages} {6015}}\BibitemShut {NoStop}%
\bibitem [{\citenamefont {Cuthbert}\ \emph {et~al.}(2007)\citenamefont
  {Cuthbert}, \citenamefont {Greedan}, \citenamefont {Vargas-Baca},
  \citenamefont {Derakhshan},\ and\ \citenamefont {Swainson}}]{ic701011r}%
  \BibitemOpen
  \bibfield  {author} {\bibinfo {author} {\bibfnamefont {H.~L.}\ \bibnamefont
  {Cuthbert}}, \bibinfo {author} {\bibfnamefont {J.~E.}\ \bibnamefont
  {Greedan}}, \bibinfo {author} {\bibfnamefont {I.}~\bibnamefont
  {Vargas-Baca}}, \bibinfo {author} {\bibfnamefont {S.}~\bibnamefont
  {Derakhshan}},\ and\ \bibinfo {author} {\bibfnamefont {I.~P.}\ \bibnamefont
  {Swainson}},\ }\href {https://doi.org/10.1021/ic701011r} {\bibfield
  {journal} {\bibinfo  {journal} {Inorg. Chem.}\ }\textbf {\bibinfo {volume}
  {46}},\ \bibinfo {pages} {8739} (\bibinfo {year} {2007})}\BibitemShut
  {NoStop}%
\bibitem [{\citenamefont {Haraguchi}\ \emph {et~al.}(2017)\citenamefont
  {Haraguchi}, \citenamefont {Michioka}, \citenamefont {Ishikawa},
  \citenamefont {Nakano}, \citenamefont {Yamochi}, \citenamefont {Ueda},\ and\
  \citenamefont {Yoshimura}}]{haraguchi2017}%
  \BibitemOpen
  \bibfield  {author} {\bibinfo {author} {\bibfnamefont {Y.}~\bibnamefont
  {Haraguchi}}, \bibinfo {author} {\bibfnamefont {C.}~\bibnamefont {Michioka}},
  \bibinfo {author} {\bibfnamefont {M.}~\bibnamefont {Ishikawa}}, \bibinfo
  {author} {\bibfnamefont {Y.}~\bibnamefont {Nakano}}, \bibinfo {author}
  {\bibfnamefont {H.}~\bibnamefont {Yamochi}}, \bibinfo {author} {\bibfnamefont
  {H.}~\bibnamefont {Ueda}},\ and\ \bibinfo {author} {\bibfnamefont
  {K.}~\bibnamefont {Yoshimura}},\ }\href
  {https://doi.org/10.1021/acs.inorgchem.6b03028} {\bibfield  {journal}
  {\bibinfo  {journal} {Inorg. Chem.}\ }\textbf {\bibinfo {volume} {56}},\
  \bibinfo {pages} {3483} (\bibinfo {year} {2017})}\BibitemShut {NoStop}%
\bibitem [{\citenamefont {Duan}\ \emph {et~al.}(2025)\citenamefont {Duan},
  \citenamefont {Jia}, \citenamefont {Fan}, \citenamefont {Ma}, \citenamefont
  {Meng}, \citenamefont {Huang},\ and\ \citenamefont {Ma}}]{cpl_42_9_090712}%
  \BibitemOpen
  \bibfield  {author} {\bibinfo {author} {\bibfnamefont {Q.}~\bibnamefont
  {Duan}}, \bibinfo {author} {\bibfnamefont {Z.}~\bibnamefont {Jia}}, \bibinfo
  {author} {\bibfnamefont {Z.}~\bibnamefont {Fan}}, \bibinfo {author}
  {\bibfnamefont {R.}~\bibnamefont {Ma}}, \bibinfo {author} {\bibfnamefont
  {J.}~\bibnamefont {Meng}}, \bibinfo {author} {\bibfnamefont {B.}~\bibnamefont
  {Huang}},\ and\ \bibinfo {author} {\bibfnamefont {T.}~\bibnamefont {Ma}},\
  }\href {http://cpl.iphy.ac.cn/en/article/doi/10.1088/0256-307X/42/9/090712}
  {\bibfield  {journal} {\bibinfo  {journal} {Chin. Phys. Lett.}\ }\textbf
  {\bibinfo {volume} {42}},\ \bibinfo {pages} {090712} (\bibinfo {year}
  {2025})}\BibitemShut {NoStop}%
\bibitem [{\citenamefont {Ashcroft}(2004)}]{PhysRevLett.92.187002}%
  \BibitemOpen
  \bibfield  {author} {\bibinfo {author} {\bibfnamefont {N.~W.}\ \bibnamefont
  {Ashcroft}},\ }\href {https://link.aps.org/doi/10.1103/PhysRevLett.92.187002}
  {\bibfield  {journal} {\bibinfo  {journal} {Phys. Rev. Lett.}\ }\textbf
  {\bibinfo {volume} {92}},\ \bibinfo {pages} {187002} (\bibinfo {year}
  {2004})}\BibitemShut {NoStop}%
\bibitem [{\citenamefont {Kresse}\ and\ \citenamefont {$\rm
  Furthm\ddot{u}ller$}(1996{\natexlab{a}})}]{PhysRevB.54.11169}%
  \BibitemOpen
  \bibfield  {author} {\bibinfo {author} {\bibfnamefont {G.}~\bibnamefont
  {Kresse}}\ and\ \bibinfo {author} {\bibfnamefont {J.}~\bibnamefont {$\rm
  Furthm\ddot{u}ller$}},\ }\href {https://doi.org/10.1103/PhysRevB.54.11169}
  {\bibfield  {journal} {\bibinfo  {journal} {Phys. Rev. B}\ }\textbf {\bibinfo
  {volume} {54}},\ \bibinfo {pages} {11169} (\bibinfo {year}
  {1996}{\natexlab{a}})}\BibitemShut {NoStop}%
\bibitem [{\citenamefont {Kresse}\ and\ \citenamefont {$\rm
  Furthm\ddot{u}ller$}(1996{\natexlab{b}})}]{CMS.0256}%
  \BibitemOpen
  \bibfield  {author} {\bibinfo {author} {\bibfnamefont {G.}~\bibnamefont
  {Kresse}}\ and\ \bibinfo {author} {\bibfnamefont {J.}~\bibnamefont {$\rm
  Furthm\ddot{u}ller$}},\ }\href {https://doi.org/10.1016/0927-0256(96)00008-0}
  {\bibfield  {journal} {\bibinfo  {journal} {Comput. Mater. Sci.}\ }\textbf
  {\bibinfo {volume} {6}},\ \bibinfo {pages} {15} (\bibinfo {year}
  {1996}{\natexlab{b}})}\BibitemShut {NoStop}%
\bibitem [{\citenamefont {Perdew}\ \emph {et~al.}(1996)\citenamefont {Perdew},
  \citenamefont {Burke},\ and\ \citenamefont
  {Ernzerhof}}]{PhysRevLett.77.3865}%
  \BibitemOpen
  \bibfield  {author} {\bibinfo {author} {\bibfnamefont {J.~P.}\ \bibnamefont
  {Perdew}}, \bibinfo {author} {\bibfnamefont {K.}~\bibnamefont {Burke}},\ and\
  \bibinfo {author} {\bibfnamefont {M.}~\bibnamefont {Ernzerhof}},\ }\href
  {https://link.aps.org/doi/10.1103/PhysRevLett.77.3865} {\bibfield  {journal}
  {\bibinfo  {journal} {Phys. Rev. Lett.}\ }\textbf {\bibinfo {volume} {77}},\
  \bibinfo {pages} {3865} (\bibinfo {year} {1996})}\BibitemShut {NoStop}%
\bibitem [{\citenamefont {Heyd}\ and\ \citenamefont
  {Scuseria}(2004)}]{10.1063/1.1760074}%
  \BibitemOpen
  \bibfield  {author} {\bibinfo {author} {\bibfnamefont {J.}~\bibnamefont
  {Heyd}}\ and\ \bibinfo {author} {\bibfnamefont {G.~E.}\ \bibnamefont
  {Scuseria}},\ }\href {https://doi.org/10.1063/1.1760074} {\bibfield
  {journal} {\bibinfo  {journal} {J. Chem. Phys.}\ }\textbf {\bibinfo {volume}
  {121}},\ \bibinfo {pages} {1187} (\bibinfo {year} {2004})}\BibitemShut
  {NoStop}%
\bibitem [{\citenamefont {Paier}\ \emph {et~al.}(2006)\citenamefont {Paier},
  \citenamefont {Marsman}, \citenamefont {Hummer}, \citenamefont {Kresse},
  \citenamefont {Gerber},\ and\ \citenamefont
  {{\'A}ngy{\'a}n}}]{paier2006screened}%
  \BibitemOpen
  \bibfield  {author} {\bibinfo {author} {\bibfnamefont {J.}~\bibnamefont
  {Paier}}, \bibinfo {author} {\bibfnamefont {M.}~\bibnamefont {Marsman}},
  \bibinfo {author} {\bibfnamefont {K.}~\bibnamefont {Hummer}}, \bibinfo
  {author} {\bibfnamefont {G.}~\bibnamefont {Kresse}}, \bibinfo {author}
  {\bibfnamefont {I.~C.}\ \bibnamefont {Gerber}},\ and\ \bibinfo {author}
  {\bibfnamefont {J.~G.}\ \bibnamefont {{\'A}ngy{\'a}n}},\ }\href
  {https://doi.org/10.1063/1.2187006} {\bibfield  {journal} {\bibinfo
  {journal} {J. Chem. Phys.}\ }\textbf {\bibinfo {volume} {124}},\ \bibinfo
  {pages} {154709} (\bibinfo {year} {2006})}\BibitemShut {NoStop}%
\bibitem [{\citenamefont {Dudarev}\ \emph {et~al.}(1998)\citenamefont
  {Dudarev}, \citenamefont {Botton}, \citenamefont {Savrasov}, \citenamefont
  {Humphreys},\ and\ \citenamefont {Sutton}}]{PhysRevB.57.1505}%
  \BibitemOpen
  \bibfield  {author} {\bibinfo {author} {\bibfnamefont {S.~L.}\ \bibnamefont
  {Dudarev}}, \bibinfo {author} {\bibfnamefont {G.~A.}\ \bibnamefont {Botton}},
  \bibinfo {author} {\bibfnamefont {S.~Y.}\ \bibnamefont {Savrasov}}, \bibinfo
  {author} {\bibfnamefont {C.~J.}\ \bibnamefont {Humphreys}},\ and\ \bibinfo
  {author} {\bibfnamefont {A.~P.}\ \bibnamefont {Sutton}},\ }\href
  {https://link.aps.org/doi/10.1103/PhysRevB.57.1505} {\bibfield  {journal}
  {\bibinfo  {journal} {Phys. Rev. B}\ }\textbf {\bibinfo {volume} {57}},\
  \bibinfo {pages} {1505} (\bibinfo {year} {1998})}\BibitemShut {NoStop}%
\bibitem [{\citenamefont {Cococcioni}\ and\ \citenamefont
  {de~Gironcoli}(2005)}]{PhysRevB.71.035105}%
  \BibitemOpen
  \bibfield  {author} {\bibinfo {author} {\bibfnamefont {M.}~\bibnamefont
  {Cococcioni}}\ and\ \bibinfo {author} {\bibfnamefont {S.}~\bibnamefont
  {de~Gironcoli}},\ }\href
  {https://link.aps.org/doi/10.1103/PhysRevB.71.035105} {\bibfield  {journal}
  {\bibinfo  {journal} {Phys. Rev. B}\ }\textbf {\bibinfo {volume} {71}},\
  \bibinfo {pages} {035105} (\bibinfo {year} {2005})}\BibitemShut {NoStop}%
\bibitem [{\citenamefont {Madsen}\ \emph {et~al.}(2018)\citenamefont {Madsen},
  \citenamefont {Carrete},\ and\ \citenamefont {Verstraete}}]{MADSEN2018140}%
  \BibitemOpen
  \bibfield  {author} {\bibinfo {author} {\bibfnamefont {G.~K.}\ \bibnamefont
  {Madsen}}, \bibinfo {author} {\bibfnamefont {J.}~\bibnamefont {Carrete}},\
  and\ \bibinfo {author} {\bibfnamefont {M.~J.}\ \bibnamefont {Verstraete}},\
  }\href {https://www.sciencedirect.com/science/article/pii/S0010465518301632}
  {\bibfield  {journal} {\bibinfo  {journal} {Comput. Phys. Commun.}\ }\textbf
  {\bibinfo {volume} {231}},\ \bibinfo {pages} {140} (\bibinfo {year}
  {2018})}\BibitemShut {NoStop}%
\bibitem [{\citenamefont {Murad}\ \emph {et~al.}(2025)\citenamefont {Murad},
  \citenamefont {Ali},\ and\ \citenamefont {Mehmood}}]{murad2025dft}%
  \BibitemOpen
  \bibfield  {author} {\bibinfo {author} {\bibfnamefont {M.}~\bibnamefont
  {Murad}}, \bibinfo {author} {\bibfnamefont {Z.}~\bibnamefont {Ali}},\ and\
  \bibinfo {author} {\bibfnamefont {S.}~\bibnamefont {Mehmood}},\ }\href
  {https://doi.org/10.1007/s00894-025-06286-y} {\bibfield  {journal} {\bibinfo
  {journal} {J. Mol. Model.}\ }\textbf {\bibinfo {volume} {31}},\ \bibinfo
  {pages} {65} (\bibinfo {year} {2025})}\BibitemShut {NoStop}%
\bibitem [{\citenamefont {Mahmoudi~Gahrouei}\ \emph {et~al.}(2024)\citenamefont
  {Mahmoudi~Gahrouei}, \citenamefont {Vlastos}, \citenamefont {D’Souza},
  \citenamefont {Odogwu},\ and\ \citenamefont
  {de~Sousa~Oliveira}}]{acs.jctc.3c01405}%
  \BibitemOpen
  \bibfield  {author} {\bibinfo {author} {\bibfnamefont {M.}~\bibnamefont
  {Mahmoudi~Gahrouei}}, \bibinfo {author} {\bibfnamefont {N.}~\bibnamefont
  {Vlastos}}, \bibinfo {author} {\bibfnamefont {R.}~\bibnamefont {D’Souza}},
  \bibinfo {author} {\bibfnamefont {E.~C.}\ \bibnamefont {Odogwu}},\ and\
  \bibinfo {author} {\bibfnamefont {L.}~\bibnamefont {de~Sousa~Oliveira}},\
  }\href {https://doi.org/10.1021/acs.jctc.3c01405} {\bibfield  {journal}
  {\bibinfo  {journal} {J. Chem. Theory Comput.}\ }\textbf {\bibinfo {volume}
  {20}},\ \bibinfo {pages} {3976} (\bibinfo {year} {2024})}\BibitemShut
  {NoStop}%
\bibitem [{\citenamefont {Long}\ \emph {et~al.}(2025)\citenamefont {Long},
  \citenamefont {Huy}, \citenamefont {Mishra},\ and\ \citenamefont
  {Makov}}]{D5RA01965F}%
  \BibitemOpen
  \bibfield  {author} {\bibinfo {author} {\bibfnamefont {N.~T.}\ \bibnamefont
  {Long}}, \bibinfo {author} {\bibfnamefont {H.~A.}\ \bibnamefont {Huy}},
  \bibinfo {author} {\bibfnamefont {N.}~\bibnamefont {Mishra}},\ and\ \bibinfo
  {author} {\bibfnamefont {G.}~\bibnamefont {Makov}},\ }\href
  {http://dx.doi.org/10.1039/D5RA01965F} {\bibfield  {journal} {\bibinfo
  {journal} {RSC Adv.}\ }\textbf {\bibinfo {volume} {15}},\ \bibinfo {pages}
  {16358} (\bibinfo {year} {2025})}\BibitemShut {NoStop}%
\bibitem [{\citenamefont {Patel}\ \emph {et~al.}(2023)\citenamefont {Patel},
  \citenamefont {Dabhi},\ and\ \citenamefont {Vora}}]{acsomega.3c06221}%
  \BibitemOpen
  \bibfield  {author} {\bibinfo {author} {\bibfnamefont {H.~S.}\ \bibnamefont
  {Patel}}, \bibinfo {author} {\bibfnamefont {V.~A.}\ \bibnamefont {Dabhi}},\
  and\ \bibinfo {author} {\bibfnamefont {A.~M.}\ \bibnamefont {Vora}},\ }\href
  {https://doi.org/10.1021/acsomega.3c06221} {\bibfield  {journal} {\bibinfo
  {journal} {ACS Omega}\ }\textbf {\bibinfo {volume} {8}},\ \bibinfo {pages}
  {43008} (\bibinfo {year} {2023})}\BibitemShut {NoStop}%
\bibitem [{\citenamefont {Wang}\ \emph
  {et~al.}(2021{\natexlab{b}})\citenamefont {Wang}, \citenamefont {Xu},
  \citenamefont {Liu}, \citenamefont {Tang},\ and\ \citenamefont
  {Geng}}]{VASPKIT}%
  \BibitemOpen
  \bibfield  {author} {\bibinfo {author} {\bibfnamefont {V.}~\bibnamefont
  {Wang}}, \bibinfo {author} {\bibfnamefont {N.}~\bibnamefont {Xu}}, \bibinfo
  {author} {\bibfnamefont {J.-C.}\ \bibnamefont {Liu}}, \bibinfo {author}
  {\bibfnamefont {G.}~\bibnamefont {Tang}},\ and\ \bibinfo {author}
  {\bibfnamefont {W.-T.}\ \bibnamefont {Geng}},\ }\href
  {https://www.sciencedirect.com/science/article/pii/S0010465521001454}
  {\bibfield  {journal} {\bibinfo  {journal} {Comput. Phys. Commun.}\ }\textbf
  {\bibinfo {volume} {267}},\ \bibinfo {pages} {108033} (\bibinfo {year}
  {2021}{\natexlab{b}})}\BibitemShut {NoStop}%
\bibitem [{\citenamefont {Gandi}\ \emph {et~al.}(2016)\citenamefont {Gandi},
  \citenamefont {Alshareef},\ and\ \citenamefont
  {Schwingenschl\"ogl}}]{Gandi2016}%
  \BibitemOpen
  \bibfield  {author} {\bibinfo {author} {\bibfnamefont {A.~N.}\ \bibnamefont
  {Gandi}}, \bibinfo {author} {\bibfnamefont {H.~N.}\ \bibnamefont
  {Alshareef}},\ and\ \bibinfo {author} {\bibfnamefont {U.}~\bibnamefont
  {Schwingenschl\"ogl}},\ }\href
  {https://doi.org/10.1021/acs.chemmater.5b04257} {\bibfield  {journal}
  {\bibinfo  {journal} {Chem. Mater.}\ }\textbf {\bibinfo {volume} {28}},\
  \bibinfo {pages} {1647} (\bibinfo {year} {2016})}\BibitemShut {NoStop}%
\bibitem [{\citenamefont {Kumar}\ and\ \citenamefont
  {Schwingenschl\"ogl}(2016)}]{PhysRevB.94.035405}%
  \BibitemOpen
  \bibfield  {author} {\bibinfo {author} {\bibfnamefont {S.}~\bibnamefont
  {Kumar}}\ and\ \bibinfo {author} {\bibfnamefont {U.}~\bibnamefont
  {Schwingenschl\"ogl}},\ }\href
  {https://link.aps.org/doi/10.1103/PhysRevB.94.035405} {\bibfield  {journal}
  {\bibinfo  {journal} {Phys. Rev. B}\ }\textbf {\bibinfo {volume} {94}},\
  \bibinfo {pages} {035405} (\bibinfo {year} {2016})}\BibitemShut {NoStop}%
\bibitem [{\citenamefont {Wang}\ \emph {et~al.}(2015)\citenamefont {Wang},
  \citenamefont {Zhang}, \citenamefont {Yu},\ and\ \citenamefont
  {Wang}}]{C5NR03813H}%
  \BibitemOpen
  \bibfield  {author} {\bibinfo {author} {\bibfnamefont {F.~Q.}\ \bibnamefont
  {Wang}}, \bibinfo {author} {\bibfnamefont {S.}~\bibnamefont {Zhang}},
  \bibinfo {author} {\bibfnamefont {J.}~\bibnamefont {Yu}},\ and\ \bibinfo
  {author} {\bibfnamefont {Q.}~\bibnamefont {Wang}},\ }\href
  {http://dx.doi.org/10.1039/C5NR03813H} {\bibfield  {journal} {\bibinfo
  {journal} {Nanoscale}\ }\textbf {\bibinfo {volume} {7}},\ \bibinfo {pages}
  {15962} (\bibinfo {year} {2015})}\BibitemShut {NoStop}%
\bibitem [{\citenamefont {Jeff}\ \emph {et~al.}(2023)\citenamefont {Jeff},
  \citenamefont {Gonzalez}, \citenamefont {Harrison}, \citenamefont {Zhao},
  \citenamefont {Fernando}, \citenamefont {Regmi}, \citenamefont {Liu},
  \citenamefont {Gutierrez}, \citenamefont {Neupane}, \citenamefont {Yang}
  \emph {et~al.}}]{Jeff_2023}%
  \BibitemOpen
  \bibfield  {author} {\bibinfo {author} {\bibfnamefont {D.~A.}\ \bibnamefont
  {Jeff}}, \bibinfo {author} {\bibfnamefont {F.}~\bibnamefont {Gonzalez}},
  \bibinfo {author} {\bibfnamefont {K.}~\bibnamefont {Harrison}}, \bibinfo
  {author} {\bibfnamefont {Y.}~\bibnamefont {Zhao}}, \bibinfo {author}
  {\bibfnamefont {T.}~\bibnamefont {Fernando}}, \bibinfo {author}
  {\bibfnamefont {S.}~\bibnamefont {Regmi}}, \bibinfo {author} {\bibfnamefont
  {Z.}~\bibnamefont {Liu}}, \bibinfo {author} {\bibfnamefont {H.~R.}\
  \bibnamefont {Gutierrez}}, \bibinfo {author} {\bibfnamefont {M.}~\bibnamefont
  {Neupane}}, \bibinfo {author} {\bibfnamefont {J.}~\bibnamefont {Yang}}, \emph
  {et~al.},\ }\href {https://doi.org/10.1088/2053-1583/acfa10} {\bibfield
  {journal} {\bibinfo  {journal} {2D Mater.}\ }\textbf {\bibinfo {volume}
  {10}},\ \bibinfo {pages} {045030} (\bibinfo {year} {2023})}\BibitemShut
  {NoStop}%
\bibitem [{\citenamefont {Brito}\ \emph {et~al.}(2017)\citenamefont {Brito},
  \citenamefont {Aguiar}, \citenamefont {Haule},\ and\ \citenamefont
  {Kotliar}}]{PhysRevB.96.195102}%
  \BibitemOpen
  \bibfield  {author} {\bibinfo {author} {\bibfnamefont {W.~H.}\ \bibnamefont
  {Brito}}, \bibinfo {author} {\bibfnamefont {M.~C.~O.}\ \bibnamefont
  {Aguiar}}, \bibinfo {author} {\bibfnamefont {K.}~\bibnamefont {Haule}},\ and\
  \bibinfo {author} {\bibfnamefont {G.}~\bibnamefont {Kotliar}},\ }\href
  {https://link.aps.org/doi/10.1103/PhysRevB.96.195102} {\bibfield  {journal}
  {\bibinfo  {journal} {Phys. Rev. B}\ }\textbf {\bibinfo {volume} {96}},\
  \bibinfo {pages} {195102} (\bibinfo {year} {2017})}\BibitemShut {NoStop}%
\bibitem [{\citenamefont {Zhou}\ \emph {et~al.}(2023)\citenamefont {Zhou},
  \citenamefont {Li}, \citenamefont {Liu}, \citenamefont {Hao}, \citenamefont
  {Dai}, \citenamefont {Wang}, \citenamefont {Yao},\ and\ \citenamefont
  {Wen}}]{PhysRevB.107.125124}%
  \BibitemOpen
  \bibfield  {author} {\bibinfo {author} {\bibfnamefont {X.}~\bibnamefont
  {Zhou}}, \bibinfo {author} {\bibfnamefont {Y.}~\bibnamefont {Li}}, \bibinfo
  {author} {\bibfnamefont {Z.}~\bibnamefont {Liu}}, \bibinfo {author}
  {\bibfnamefont {J.}~\bibnamefont {Hao}}, \bibinfo {author} {\bibfnamefont
  {Y.}~\bibnamefont {Dai}}, \bibinfo {author} {\bibfnamefont {Z.}~\bibnamefont
  {Wang}}, \bibinfo {author} {\bibfnamefont {Y.}~\bibnamefont {Yao}},\ and\
  \bibinfo {author} {\bibfnamefont {H.-H.}\ \bibnamefont {Wen}},\ }\href
  {https://link.aps.org/doi/10.1103/PhysRevB.107.125124} {\bibfield  {journal}
  {\bibinfo  {journal} {Phys. Rev. B}\ }\textbf {\bibinfo {volume} {107}},\
  \bibinfo {pages} {125124} (\bibinfo {year} {2023})}\BibitemShut {NoStop}%
\bibitem [{\citenamefont {Yin}\ \emph {et~al.}(2014)\citenamefont {Yin},
  \citenamefont {Shi},\ and\ \citenamefont {Yan}}]{adma.201306281}%
  \BibitemOpen
  \bibfield  {author} {\bibinfo {author} {\bibfnamefont {W.-J.}\ \bibnamefont
  {Yin}}, \bibinfo {author} {\bibfnamefont {T.}~\bibnamefont {Shi}},\ and\
  \bibinfo {author} {\bibfnamefont {Y.}~\bibnamefont {Yan}},\ }\href
  {https://advanced.onlinelibrary.wiley.com/doi/abs/10.1002/adma.201306281}
  {\bibfield  {journal} {\bibinfo  {journal} {Adv. Mater.}\ }\textbf {\bibinfo
  {volume} {26}},\ \bibinfo {pages} {4653} (\bibinfo {year}
  {2014})}\BibitemShut {NoStop}%
\bibitem [{\citenamefont {Orain}\ \emph {et~al.}(2017)\citenamefont {Orain},
  \citenamefont {Bernu}, \citenamefont {Mendels}, \citenamefont {Clark},
  \citenamefont {Aidoudi}, \citenamefont {Lightfoot}, \citenamefont {Morris},\
  and\ \citenamefont {Bert}}]{PhysRevLett.118.237203}%
  \BibitemOpen
  \bibfield  {author} {\bibinfo {author} {\bibfnamefont {J.-C.}\ \bibnamefont
  {Orain}}, \bibinfo {author} {\bibfnamefont {B.}~\bibnamefont {Bernu}},
  \bibinfo {author} {\bibfnamefont {P.}~\bibnamefont {Mendels}}, \bibinfo
  {author} {\bibfnamefont {L.}~\bibnamefont {Clark}}, \bibinfo {author}
  {\bibfnamefont {F.~H.}\ \bibnamefont {Aidoudi}}, \bibinfo {author}
  {\bibfnamefont {P.}~\bibnamefont {Lightfoot}}, \bibinfo {author}
  {\bibfnamefont {R.~E.}\ \bibnamefont {Morris}},\ and\ \bibinfo {author}
  {\bibfnamefont {F.}~\bibnamefont {Bert}},\ }\href
  {https://link.aps.org/doi/10.1103/PhysRevLett.118.237203} {\bibfield
  {journal} {\bibinfo  {journal} {Phys. Rev. Lett.}\ }\textbf {\bibinfo
  {volume} {118}},\ \bibinfo {pages} {237203} (\bibinfo {year}
  {2017})}\BibitemShut {NoStop}%
\bibitem [{\citenamefont {Cordero}\ \emph {et~al.}(2008)\citenamefont
  {Cordero}, \citenamefont {G{\'o}mez}, \citenamefont {Platero-Prats},
  \citenamefont {Rev{\'e}s}, \citenamefont {Echeverr{\'i}a}, \citenamefont
  {Cremades}, \citenamefont {Barrag{\'a}n},\ and\ \citenamefont
  {Alvarez}}]{B801115J}%
  \BibitemOpen
  \bibfield  {author} {\bibinfo {author} {\bibfnamefont {B.}~\bibnamefont
  {Cordero}}, \bibinfo {author} {\bibfnamefont {V.}~\bibnamefont {G{\'o}mez}},
  \bibinfo {author} {\bibfnamefont {A.~E.}\ \bibnamefont {Platero-Prats}},
  \bibinfo {author} {\bibfnamefont {M.}~\bibnamefont {Rev{\'e}s}}, \bibinfo
  {author} {\bibfnamefont {J.}~\bibnamefont {Echeverr{\'i}a}}, \bibinfo
  {author} {\bibfnamefont {E.}~\bibnamefont {Cremades}}, \bibinfo {author}
  {\bibfnamefont {F.}~\bibnamefont {Barrag{\'a}n}},\ and\ \bibinfo {author}
  {\bibfnamefont {S.}~\bibnamefont {Alvarez}},\ }\href
  {http://dx.doi.org/10.1039/B801115J} {\bibfield  {journal} {\bibinfo
  {journal} {Dalton Trans.}\ ,\ \bibinfo {pages} {2832}} (\bibinfo {year}
  {2008})}\BibitemShut {NoStop}%
\bibitem [{\citenamefont {Becke}\ and\ \citenamefont
  {Edgecombe}(1990)}]{10.1063/1.458517}%
  \BibitemOpen
  \bibfield  {author} {\bibinfo {author} {\bibfnamefont {A.~D.}\ \bibnamefont
  {Becke}}\ and\ \bibinfo {author} {\bibfnamefont {K.~E.}\ \bibnamefont
  {Edgecombe}},\ }\href {https://doi.org/10.1063/1.458517} {\bibfield
  {journal} {\bibinfo  {journal} {J. Chem. Phys.}\ }\textbf {\bibinfo {volume}
  {92}},\ \bibinfo {pages} {5397} (\bibinfo {year} {1990})}\BibitemShut
  {NoStop}%
\bibitem [{\citenamefont {Silvi}\ and\ \citenamefont
  {Savin}(1994)}]{Silvi1994}%
  \BibitemOpen
  \bibfield  {author} {\bibinfo {author} {\bibfnamefont {B.}~\bibnamefont
  {Silvi}}\ and\ \bibinfo {author} {\bibfnamefont {A.}~\bibnamefont {Savin}},\
  }\href {https://doi.org/10.1038/371683a0} {\bibfield  {journal} {\bibinfo
  {journal} {Nature}\ }\textbf {\bibinfo {volume} {371}},\ \bibinfo {pages}
  {683} (\bibinfo {year} {1994})}\BibitemShut {NoStop}%
\end{thebibliography}%
\bibliographystyle{style}

\end{document}